\documentclass[structabstract]{aa}
\usepackage{graphicx}
\usepackage{txfonts}
\usepackage{natbib}
\usepackage{color}
\begin{document}
\title{Radial velocities for the Hipparcos--Gaia Hundred-Thousand-Proper-Motion project}

\author{J.H.J. de Bruijne
  \and
  A.--C. Eilers
}

\institute{Research and Scientific Support Department in the Directorate of Science and Robotic Exploration of the European Space Agency, Postbus 299, NL--2200AG, Noordwijk, The Netherlands\\
  \email{jdbruijn@rssd.esa.int}
}


\color{black}

\begin{abstract}
{The Hundred-Thousand-Proper-Motion (HTPM) project will determine the proper motions of $\sim$$113,500$ stars using a $\sim$23-year baseline. The proper motions will be based on space-based measurements exclusively, with the Hipparcos data, with epoch $1991.25$, as first epoch and with the first intermediate-release Gaia astrometry, with epoch $\sim$$2014.5$, as second epoch. The expected HTPM proper-motion standard errors are $30$--$190$~$\mu$as~yr$^{-1}$, depending on stellar magnitude.}
{Depending on the astrometric characteristics of an object, in particular its distance and velocity, its radial velocity can have a significant impact on the determination of its proper motion. The impact of this perspective acceleration is largest for fast-moving, nearby stars. Our goal is to determine, for each star in the Hipparcos catalogue, the radial-velocity standard error that is required to guarantee a negligible contribution of perspective acceleration to the HTPM proper-motion precision.}
{We employ two evaluation criteria, both based on Monte-Carlo simulations, with which we determine which stars need to be spectroscopically (re-)measured. Both criteria take the Hipparcos measurement errors into account. The first criterion, the Gaussian criterion, is applicable to nearby stars. For distant stars, this criterion works but returns overly pessimistic results. We therefore use a second criterion, the robust criterion, which is equivalent to the Gaussian criterion for nearby stars but avoids biases for distant stars and/or objects without literature radial velocity. The robust criterion is hence our prefered choice for all stars, regardless of distance.}
{For each star in the Hipparcos catalogue, we determine the confidence level with which the available radial velocity and its standard error, taken from the XHIP compilation catalogue, are acceptable. We find that for \color{black} $97$ \color{black} stars, the radial velocities available in the literature are insufficiently precise for a $68.27\%$ confidence level. If requiring this level to be $95.45\%$, or even $99.73\%$, the number of stars increases to \color{black} $247$ \color{black} or \color{black} $382$\color{black}, respectively. We also identify \color{black} $109$ \color{black} stars for which radial velocities are currently unknown yet need to be acquired to meet the $68.27\%$ confidence level. For higher confidence levels ($95.45\%$ or $99.73\%$), the number of such stars increases to \color{black} $1,071$ \color{black} or \color{black} $6,180$\color{black}, respectively.}
{To satisfy the radial-velocity requirements coming from our study will be a daunting task consuming a significant amount of spectroscopic telescope time. The required radial-velocity measurement precisions vary from source to source. Typically, they are modest, below $25$~km~s$^{-1}$, but they can be as stringent as $0.04$~km~s$^{-1}$ for individual objects like Barnard's star. Fortunately, the follow-up spectroscopy is not time-critical since the HTPM proper motions can be corrected {\it {a posteriori}} {\rm{once (improved) radial velocities become available.}}}
\end{abstract}


\maketitle

\color{black}

\section{Introduction}

Gaia \citep[e.g.,][]{2001A&A...369..339P,2008IAUS..248..217L} is the upcoming astrometry mission of the European Space Agency (ESA), following up on the success of the Hipparcos mission \citep{1997ESASP1200.....P,1997A&A...323L..49P,2009aaat.book.....P}. Gaia's science objective is to unravel the kinematical, dynamical, and chemical structure and evolution of our galaxy, the Milky Way \citep[e.g.,][]{2010MNRAS.408..935G}. In addition, Gaia's data will revolutionise many other areas of (astro)physics, e.g., stellar structure and evolution, stellar variability, double and multiple stars, solar-system bodies, fundamental physics, and exo-planets \citep[e.g.,][]{2008IAUS..248...59P,2008P&SS...56.1812T,2010IAUS..261..306M,2011EAS....45..161E,2011EAS....45..273S,2011PhRvD..84l2001M}. During its five-year lifetime, Gaia will survey the full sky and repeatedly observe the brightest 1,000 million objects, down to $20^{\rm th}$ magnitude \citep[e.g.,][]{2010SPIE.7731E..35D}. Gaia's science data comprises absolute astrometry, broad-band photometry, and low-resolution spectro-photometry. Medium-resolution spectroscopic data will be obtained for the brightest 150 million sources, down to $17^{\rm th}$ magnitude. The final Gaia catalogue, due in $\sim$$2021$, will contain astrometry (positions, parallaxes, and proper motions) with standard errors less than $10$~micro-arcsecond ($\mu$as, $\mu$as~yr$^{-1}$ for proper motions) for stars brighter than $12^{\rm th}$ magnitude, $25~\mu$as for stars at $15^{\rm th}$ magnitude, and $300~\mu$as at magnitude 20 \citep{2012Ap&SS.tmp...68D}. Milli-magnitude-precision photometry \citep{2010A&A...523A..48J} allows to get a handle on effective temperature, surface gravity, metallicity, and reddening of all stars \citep{2010MNRAS.403...96B}. The spectroscopic data allows the determination of radial velocities with errors of $1~{\rm km~s}^{-1}$ at the bright end and $15~{\rm km~s}^{-1}$ at magnitude 17 \citep{2005MNRAS.359.1306W,2011EAS....45..189K} as well as astrophysical diagnostics such as effective temperature and metallicity for the brightest few million objects \citep{2011A&A...535A.106K}. Clearly, these performances will only be reached with a total of five years of collected data and only after careful calibration.

Intermediate releases of the data -- obviously with lower quality and/or reduced contents compared to the final catalogue -- are planned, the first one around two years after launch, which is currently foreseen for the second half of 2013. The Hundred-Thousand-Proper-Motion (HTPM) project \citep{FM-040}, conceived and led by Fran\c{c}ois Mignard at the Observatoire de la C\^{o}te d'Azur, is part of the first intermediate release. Its goal is to determine the absolute proper motions of the $\sim$$113,500$ brightest stars in the sky using Hipparcos astrometry for the first epoch and early Gaia astrometry for the second. Clearly, the HTPM catalogue will have a limited lifetime since it will be superseded by the final Gaia catalogue in $\sim$$2021$. Nevertheless, the HTPM is a scientifically interesting as well as unique catalogue: the $\sim$23-year temporal baseline, with a mean Hipparcos epoch of $1991.25$ and a mean Gaia epoch around $2014.5$, allows a significant improvement of the Hipparcos proper motions, which have typical precisions at the level of $1$~milli-arcsecond~yr$^{-1}$ (mas~yr$^{-1}$): the expected HTPM proper-motion standard errors\footnote
{The HTPM proper motions will be limited in precision by the Hipparcos parallax uncertainties, which are typically $\sim$1~mas (the typical HTPM proper-motion standard error is hence $1~{\rm mas} / 23~{\rm yr} \approx 40~\mu{\rm as~yr}^{-1}$). The first intermediate-release Gaia catalogue is based on just $\sim$$12$ months of data, which is generally insufficient to unambiguously lift the degeneracy between proper motion and parallax for all stars. The underlying astrometric global iterative solution \citep{2012A&A...538A..78L} will hence be based on a two- rather than five-parameter source model, fitting for position $(\alpha, \delta)$ at the mean Gaia epoch only. The Hipparcos parallax is hence needed to correct the Gaia transit observations for the parallactic effect allowing to transform apparent directions into barycentric positions.}
are $40$--$190~\mu$as~yr$^{-1}$ for the proper motion in right ascension $\mu_{\alpha^*}$ and $30$--$150$~$\mu$as~yr$^{-1}$ for the proper motion in declination $\mu_{\delta}$, primarily depending on magnitude (we use the common Hipparcos notation $\alpha^* = \alpha \cdot \cos\delta$; \citealp[Section 1.2.5]{1997ESASP1200.....P}). A clear advantage of combining astrometric data from the Hipparcos and Gaia missions is that the associated proper motions will be, by construction and IAU resolution, in the system of the International Celestial Reference System (ICRS), i.e., the proper motions will be absolute rather than relative. In this light, it is important to realise that massive, modern-day proper-motion catalogues, such as UCAC-3 \citep{2010AJ....139.2184Z}, often contain relative proper motions only and that they can suffer from substantial, regional, systematic distortions in their proper-motion systems, up to levels of 10~mas~yr$^{-1}$ or more \citep[e.g.,][]{2008A&A...488..401R,2010AJ....139.2440R,2011RAA....11.1074L}.

It is a well-known geometrical feature, for instance already described by \citeauthor{1900AN....154...65S} in \citeyear{1900AN....154...65S}, that for fast-moving, nearby stars, it is essential to know the radial velocity for a precise measurement and determination of proper motion. In fact, this so-called secular or perspective acceleration on the sky was taken into account in the determination of the Hipparcos proper motions for 21 stars \citep[Section 1.2.8]{1997ESASP1200.....P} and the same will be done for Gaia, albeit for a larger sample of nearby stars. Clearly, the inverse relationship also holds: with a precise proper motion available, a so-called astrometric radial velocity can be determined, independent of the spectroscopically measured quantity (see \citealt{2003A&A...401.1185L} for a precise definition and meaning of [astrometric] radial velocity). With this method, \citet{1999A&A...348.1040D} determined\footnote
{These authors also describe two other methods to derive astrometric radial velocities, namely by measuring changing annual parallax or by measuring changing angular extent of a moving group of stars \citep{2002A&A...381..446M}. The latter method also provides, as a bonus, improved parallaxes to moving-group members \citep[e.g.,][]{1999MNRAS.310..585D,2001A&A...367..111D}.}
the astrometric radial velocities for 17 stars, from Hipparcos proper motions combined with Astrographic Catalogue positions at earlier epochs. Although \citet{1999A&A...348.1040D} reached relatively modest astrometric-radial-velocity precisions, typically a few tens of km~s$^{-1}$, their results are interesting since they provide direct and independent constraints on various physical phenomena affecting spectroscopic radial velocities, for instance gravitational redshifts, stellar rotation, convection, and pulsation. In our study, however, we approach (astrometric) radial velocities from the other direction since our interest is to determine accurate HTPM proper motions which are not biased by unmodeled perspective effects. In other words: we aim to establish for which stars in the forthcoming HTPM catalogue the currently available (spectroscopic) radial velocity and associated standard error are sufficient to guarantee, with a certain confidence level, a negligible perspective-acceleration-induced error in the HTPM proper motion. For stars without a literature value of the radial velocity, we establish whether -- and, if yes, with what standard error -- a radial velocity needs to be acquired prior to the construction of the HTPM catalogue. Section~\ref{sec:XHIP} describes the available astrometric and spectroscopic data. The propagation model of star positions is outlined in Section~\ref{sec:propagation_model}. We investigate the influence of the radial velocity on HTPM proper motions in Section~\ref{sec:rv_influence} and develop two evaluation criteria in Section~\ref{sec:evaluation_criteria}. We employ these in Section~\ref{sec:application}. We discuss our results in Section~\ref{sec:discussion} and give our final conclusions in Section~\ref{sec:conclusion}.

\section{The XHIP catalogue}\label{sec:XHIP}

As source for the Hipparcos astrometry and literature radial velocities, we used the eXtended Hipparcos compilation catalogue (CDS catalogue V/137), also known as XHIP \citep{2012AstL...38..331A}. This catalogue complements the $117,955$ entries with astrometry in the Hipparcos catalogue with a set of $116,096$ spectral classifications, $46,392$ radial velocities, and $18,549$ iron abundances from various literature sources.

\subsection{Astrometry}\label{sec:XHIP_astrometry}

The starting point for the XHIP compilation was the new reduction of the Hipparcos data \citep[CDS catalogue I/311]{2007ASSL..350.....V,2008yCat.1311....0V}, also known as HIP-2. Realising that stars with multiple components were solved individually, rather than as systems, by \citeauthor{2007ASSL..350.....V} for the sake of expediency, \citeauthor{2012AstL...38..331A} reverted to the original HIP-1 astrometry \citep[CDS catalogue I/239]{1997ESASP1200.....P,1997yCat.1239....0E} in those cases where multiplicity is indicated and where the formal parallax \color{black} standard \color{black} error in HIP-2 exceeds that in HIP-1. This applies to $1,922$ entries. In addition, \citeauthor{2012AstL...38..331A} included the Tycho-2 catalogue \citep[CDS catalogue I/259]{2000yCat.1259....0H,2000A&A...355L..27H} in their XHIP proper-motion data. In the absence of Tycho-2 proper motions, HIP-2 proper motions were forcibly used. When multiplicity is indicated, Hipparcos proper motions were replaced by Tycho-2 values in those cases where the latter are more precise. When multiplicity is not indicated, Tycho-2 proper motions replaced Hipparcos values if the associated standard errors exceed the Tycho-2 standard errors by a factor three or more. In all other cases, a mean HIP-2 -- Tycho-2 proper motion was constructed and used, weighted by the inverse squared \color{black} standard \color{black} errors; this applies to $92,269$ entries.

\subsection{Radial velocities}\label{sec:XHIP_radial_velocities}

The XHIP catalogue contains radial velocities for $46,392$ of the $117,955$ entries, carefully compiled by \citeauthor{2012AstL...38..331A} from 47 literature sources. The vast majority of measurements have formal measurement precisions, i.e., radial-velocity standard errors ($1,753$ measurements lack \color{black} standard \color{black} errors; see Section~\ref{sec:stars_without_radial_velocity}). In addition, all radial velocities have a quality flag:
\begin{itemize}
\item An 'A' rating ($35,932$ entries) indicates that the standard errors are generally reliable.
\item A 'B' rating ($4,239$ entries) indicates potential, small, uncorrected, systematic errors.
\item A 'C' rating ($3,465$ entries) indicates larger systematic errors, while not excluding suitability for population analyses.
\item A 'D' rating ($2,756$ entries) indicates serious problems, meaning that these stars may not be suitable for statistical analyses. A 'D' rating is assigned whenever:
\begin{enumerate}
\item the radial-velocity standard error is not available,
\item the star is an un(re)solved binary,
\item the star is a Wolf-Rayet star or a white dwarf that is not a component of a resolved binary, or
\item different measurements yield inconsistent results.
\end{enumerate}
\end{itemize}

\begin{figure}[t]
  \includegraphics[width = \columnwidth]{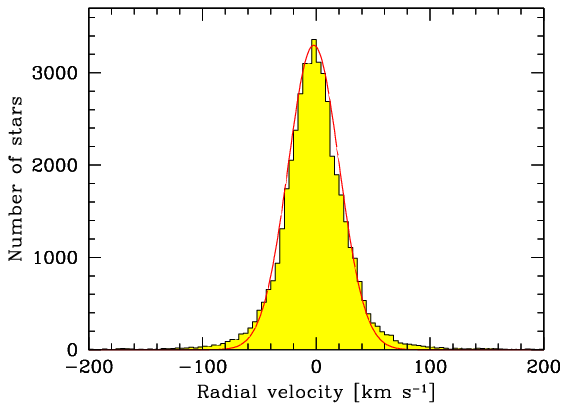}
  \caption{Distribution of all $46,392$ radial velocities contained in the XHIP catalogue. The smooth, red curve fits the histogram with a Gaussian normal distribution. The best-fit mean and standard deviation are $\mu = -2.21$~km~s$^{-1}$ and $\sigma = 22.44$~km~s$^{-1}$, respectively.}\label{fig:all_vr}
\end{figure}

The majority of stars in the XHIP catalogue ($71,563$ entries to be precise) have no measured radial velocity. All we can reasonably assume for these stars is that their radial-velocity distribution is statistically identical to the radial-velocity distribution of the entries with known radial velocities. Figure~\ref{fig:all_vr} shows this distribution. It is fairly well represented by a normal distribution with a mean $\mu = -2.21$~km~s$^{-1}$ and standard deviation $\sigma = 22.44$~km~s$^{-1}$ (the median is $-2.00$~km~s$^{-1}$). The observed distribution has low-amplitude, broad wings as well as a small number of real 'outliers', with heliocentric radial velocities up to plus-or-minus several hundred km~s$^{-1}$. A small fraction of these stars are early-type runaway\footnote{The Hipparcos catalogue does not contain hyper-velocity stars.} stars \citep{2001A&A...365...49H} but the majority represent nearby stars in the (non-rotating) halo of our galaxy. The bulk of the stars, those in the main peak, are (thin-)disc stars, co-rotating with the Sun around the galactic centre. In theory, the main peak can be understood, and also be modeled in detail and hence be used to statistically predict the radial velocities for objects without literature values, as a combination of the reflex of the solar motion with respect to the local standard of rest \citep{1965gast.conf...61D,2010MNRAS.403.1829S}, the effect of differential galactic rotation \citep{1927BAN.....3..275O}, and the random motion of stars \citep{1907..............S}. In practice, however, such a modeling effort would be massive, touching on a wide variety of (sometimes poorly understood) issues such as the asymmetric drift, the tilt and vertex deviation of the velocity ellipsoid, mixing and heating of stars as function of age, the height of the sun above the galactic plane \citep{2007MNRAS.378..768J}, the dynamical coupling of the local kinematics to the galactic bar and spiral arms \citep{2011MNRAS.418.1423A}, large-scale deviations of the local velocity field caused by the Gould Belt \citep{2006AJ....132.1052E}, migration of stars in the disc \citep{2009MNRAS.399.1145S}, etc. To model these effects, and hence be able to predict a more refined radial velocity for any star as function of its galactic coordinates, distance, and age, is clearly beyond the scope of this paper. We will come back to this issue in Section~\ref{sec:stars_without_radial_velocity}.

\section{Propagation model}\label{sec:propagation_model}

\subsection{The full model}\label{sec:full_model}

Let us denote the celestial position of a star at time $t_0$ (in years) in equatorial coordinates in radians by $(\alpha_0, \delta_0)$, its distance in parsec by $r = 1000 \cdot \varpi^{-1}$, with the parallax $\varpi$ in mas, and its proper-motion components in equatorial coordinates in mas~yr$^{-1}$ by $(\mu_{\alpha}, \mu_{\delta})$. The three-dimensional position of a star in Cartesian equatorial coordinates at time $t_0$ is then given by:
\begin{eqnarray}
x_0 & = & r \cdot \cos\alpha_0\cdot\cos\delta_0;\nonumber\\
y_0 & = & r \cdot \sin\alpha_0\cdot\cos\delta_0;\nonumber\\
z_0 & = & r \cdot \sin\delta_0.
\end{eqnarray}
With $v_r$ the star's radial velocity in km~s$^{-1}$, it is customary to define the 'radial proper motion' as $\mu_r = v_r \cdot r^{-1}$. The linear velocity in pc~yr$^{-1}$ is then given by:
\begin{eqnarray}
v_{\alpha} &=& \mu_{\alpha} \cdot B \cdot A_V \cdot \varpi^{-1};\nonumber\\
v_{\delta} &=& \mu_{\delta} \cdot B \cdot A_V \cdot \varpi^{-1};\nonumber\\
v_r       &=& \mu_r        \cdot B,
\end{eqnarray}
where $A_V = 4.740,470,446$~km~yr~s$^{-1}$ is the astronomical unit and the factor $B = A_p \cdot A_z^{-1} = 1.022,712,169 \cdot 10^{-6}$~pc~s~km$^{-1}$~yr$^{-1}$ changes km~s$^{-1}$ to pc~yr$^{-1}$ \citep[][Table 1.2.2]{1997ESASP1200.....P} so that $B \cdot A_V = 4.848,136,811 \cdot 10^{-6}$~pc~AU$^{-1}$. Transforming these equations into Cartesian coordinates leads to:
\begin{eqnarray}
v_x &=&          - v_{\alpha}\sin\alpha_0 - v_{\delta}\cos\alpha_0\sin\delta_0 + v_r\cos\alpha_0\cos\delta_0;\nonumber\\
v_y &=& \phantom{-}v_{\alpha}\cos\alpha_0 - v_{\delta}\sin\alpha_0\sin\delta_0 + v_r\sin\alpha_0\cos\delta_0;\nonumber\\
v_z &=& \phantom{-}\phantom{v_{\alpha}\sin\alpha_0} \phantom{-\ \ } v_{\delta}\phantom{\cos\alpha_0}\cos\delta_0 + v_r\phantom{\cos\alpha_0}\sin\delta_0.
\end{eqnarray}
Since the motion of stars, or the barycentre of multiple systems, is to near-perfect approximation rectilinear over time scales of a few decades, the position of a star at time $t_1$, after a time $t = t_1 - t_0$, now simply follows by applying the propagation model:
\begin{eqnarray}
x(t) & = & x_0 + v_x \cdot t;\nonumber\\
y(t) & = & y_0 + v_y \cdot t;\nonumber\\
z(t) & = & z_0 + v_z \cdot t.
\end{eqnarray}
Transforming this back into equatorial coordinates returns the celestial position $(\alpha(t), \delta(t))$ of the star at time $t$:
\begin{eqnarray}
\alpha(t)& = &\arctan\left[\frac{y(t)}{x(t)}\right];\nonumber\\
\delta(t)& = &\arctan\left[\frac{z(t)}{\sqrt{x(t)^2+y(t)^2}}\right].
\end{eqnarray}
So, in summary, it is straightforward to compute the future position of a star on the sky once the initial celestial coordinates, the proper motion, parallax, radial velocity, and time interval are given. This, however, is what nature and Gaia will do for us: the initial celestial coordinates correspond to the Hipparcos epoch ($1991.25$) and the final coordinates $(\alpha(t), \delta(t))$, with $t \approx 23$~yr, will come from the first Gaia astrometry. The HTPM project will determine the proper-motion components $(\mu_{\alpha}, \mu_{\delta})$ from known initial and final celestial coordinates for given time interval, parallax, and radial velocity. Unfortunately, it is not possible to express the proper-motion components in a closed (analytical) form as function of $(\alpha_0, \delta_0)$, $(\alpha(t), \delta(t))$, $\varpi$, $v_r$, and $t$ since the underlying set of equations is coupled. The derivation of the proper-motion components hence requires a numerical solution. We implemented this solution using Newton--Raphson iteration and refer to this solution as the 'full model'. This model, however, is relatively slow for practical implementation. We hence decided to also implement a 'truncated model' with analytical terms up to and including $t^3$ (Section~\ref{sec:truncated_model}), which is about ten times faster and sufficiently precise for our application (Section~\ref{sec:truncated_model_accuracy}).

\subsection{The truncated model}\label{sec:truncated_model}

\citet{FM-040} shows that the full model (Section~\ref{sec:full_model}) can be truncated up to and including third-order terms in time $t$ without significant loss in accuracy (Section~\ref{sec:truncated_model_accuracy}). Equations (\ref{eq:forward1}) and (\ref{eq:forward2}) below show the forward propagation for right ascension and declination, respectively. By forward propagation, we mean computing the positional displacements $\Delta\alpha$ and $\Delta\delta$ of a star for given proper motion, parallax, radial velocity, and time interval $t = t_1 - t_0$:
\begin{eqnarray}
\Delta\alpha^* = \Delta\alpha\cos\delta_0 = (\alpha(t)-\alpha_0)\cos\delta_0 = \phantom{+} \nonumber\\
\left[\mu_{\alpha}\right] \cdot t^1 + \nonumber\\
-\left[\mu_r\mu_{\alpha}-\tan\delta_0\mu_{\alpha}\mu_{\delta}\right] \cdot t^2 + \nonumber\\
\phantom{0} +\left[\mu_r^2\mu_{\alpha}-2\tan\delta_0\mu_r\mu_{\alpha}\mu_{\delta}+\tan^2\delta_0\mu_{\alpha}\mu_{\delta}^2-\frac{\mu_{\alpha}^3}{3\cos^2\delta_0}\right] \cdot t^3 + \nonumber\\
\mathcal{O}(t^4);\label{eq:forward1}
\end{eqnarray}
\begin{eqnarray}
\Delta\delta = \delta(t)-\delta_0 = \phantom{+} \nonumber\\
\left[\mu_{\delta}\right] \cdot t^1 + \nonumber\\
-\left[\mu_r\mu_{\delta}+\frac{\tan\delta_0}{2}\mu_{\alpha}^2\right] \cdot t^2 + \nonumber\\
\phantom{000000} +\left[\mu_r^2\mu_{\delta}+\tan\delta_0\mu_r\mu_{\alpha}^2-\frac{\mu_{\alpha}^2\mu_{\delta}}{2\cos^2\delta_0}-\frac{\mu_{\delta}^3}{3}\right] \cdot t^3 + \nonumber\\
\mathcal{O}(t^4).\label{eq:forward2}
\end{eqnarray}
Equations (\ref{eq:backward1}) and (\ref{eq:backward2}) below show the \color{black} backward solution \color{black} for the proper motion in right ascension and declination, respectively. By \color{black} backward solution\color{black}, we mean computing the proper-motion components $(\mu_{\alpha}, \mu_{\delta})$ from the initial and final celestial positions $(\alpha_0, \delta_0)$ and $(\alpha(t), \delta(t))$, for given parallax, radial velocity, and time interval $t = t_1 - t_0$:
\begin{eqnarray}
\mu_{\alpha} \cdot t = \phantom{+} \nonumber\\
\Delta\alpha^* + \Delta\alpha^*\mu_rt + \nonumber\\
\phantom{0000000} - \tan\delta_0\Delta\alpha^*\Delta\delta - \tan\delta_0\Delta\alpha^*\Delta\delta\mu_rt + \nonumber\\
+ \frac{3\cos^2\delta_0-1}{6\cos^2\delta_0}(\Delta\alpha^*)^3 \phantom{+}
\label{eq:backward1}
\end{eqnarray}
\begin{eqnarray}
\mu_{\delta} \cdot t = \phantom{+} \nonumber\\
\Delta\delta + \Delta\delta\mu_rt + \nonumber\\
\phantom{0000000} + \frac{1}{2}\tan\delta_0(\Delta\alpha^*)^2 + \frac{1}{2}\tan\delta_0(\Delta\alpha^*)^2\mu_r t + \nonumber\\
+ \frac{2\cos^2\delta_0-1}{2\cos^2\delta_0}(\Delta\alpha^*)^2\Delta\delta + \frac{\Delta\delta^3}{3} \phantom{+}
\label{eq:backward2}
\end{eqnarray}
It is straightforward to insert Equations (\ref{eq:forward1}) and (\ref{eq:forward2}) into Equations (\ref{eq:backward1}) and (\ref{eq:backward2}) to demonstrate that only terms of order $t^4$ and higher are left.

\subsection{Accuracy of the truncated model}\label{sec:truncated_model_accuracy}

To quantify that the truncation of the full model up to and including third-order terms in time is sufficient for the HTPM application, Table~\ref{tab:error} shows the errors in derived proper motions over an interval of 25 years induced by the truncation of the model when using the approximated Equations (\ref{eq:forward1})--(\ref{eq:backward2}) for an 'extreme' star (i.e., nearby, fast-moving and hence sensitive to perspective-acceleration effects) as function of declination. Four cases have been considered. First, Equations (\ref{eq:forward1})--(\ref{eq:forward2}) up to and including first-order terms in time were used for the forward propagation and Equations (\ref{eq:backward1})--(\ref{eq:backward2}) were used for the \color{black} backward solution\color{black}. This is indicated by the heading $\mathcal{O}(t^1)$. The difference between the proper motion used as input and the proper motion derived from Equations (\ref{eq:backward1})--(\ref{eq:backward2}) is listed in the table and can reach several mas~yr$^{-1}$ close to the celestial poles. The second case ('$\mathcal{O}(t^2)$') is similar to the first case but includes also second-order terms in time for the forward propagation. The proper-motion errors are now much reduced, by about three orders of magnitude, but can still reach $10~\mu$as~yr$^{-1}$, which is significant given the predicted HTPM standard errors ($30$--$190$~$\mu$as~yr$^{-1}$, depending on magnitude). The third case ('$\mathcal{O}(t^3)$') also includes third-order terms in time. The proper-motion errors are now negligible, reaching only up to $10$~nano-arcsecond~yr$^{-1}$. The fourth case uses the full model for the forward propagation and the Newton--Raphson iteration for the \color{black} backward solution \color{black} and recovers the input proper motions with sub-nano-arcsecond~yr$^{-1}$ errors.

\begin{table}[t]
\caption{Proper-motion errors accumulated over $t = 25$~yr as function of declination due to truncation of the full model up to and including first, second, and third-order terms in time for an 'extreme', i.e., nearby, fast-moving star: $\varpi = 500$~mas ($r = 2$~pc), $\mu_{\alpha} = \mu_{\delta} = 2000$~mas~yr$^{-1}$, and $v_r = 50$~km~s$^{-1}$. The unit nas stands for nano-arcsecond.}\label{tab:error}
\begin{center}
\begin{tabular}[h]{rrrrrrrrr}
\hline\hline
\\[-8pt]
& \multicolumn{2}{c}{\color{black}$\mathcal{O}(t^2)$\color{black}} & \multicolumn{2}{c}{\color{black}$\mathcal{O}(t^3)$\color{black}} & \multicolumn{2}{c}{\color{black}$\mathcal{O}(t^4)$\color{black}} & \multicolumn{2}{c}{Full model}\\
 $\delta\ [^{\circ}]$ & $\mu_{\alpha}$ & $\mu_{\delta}$ & $\mu_{\alpha}$ & $\mu_{\delta}$ & $\mu_{\alpha}$ & $\mu_{\delta}$ & $\mu_{\alpha}$ & $\mu_{\delta}$\\
& \multicolumn{2}{c}{[mas~yr$^{-1}$]} & \multicolumn{2}{c}{[$\mu$as~yr$^{-1}$]} & \multicolumn{2}{c}{[nas~yr$^{-1}$]} & \multicolumn{2}{c}{[nas~yr$^{-1}$]}\\
\hline
$85$&$4.3$&$-4.1$&$6.1$&$5.8$&$-5.1$&$6.7$&$0.0$&$0.0$\\
$75$&$0.5$&$-2.2$&$1.4$&$0.1$&$0.0$&$0.6$&$0.0$&$0.0$\\
$60$&$-0.4$&$-1.7$&$0.6$&$-0.2$&$0.0$&$0.1$&$0.0$&$0.0$\\
$45$&$-0.8$&$-1.5$&$0.3$&$-0.2$&$0.0$&$0.0$&$0.0$&$0.0$\\
$30$&$-1.0$&$-1.4$&$0.1$&$-0.2$&$0.0$&$0.0$&$0.0$&$0.0$\\
$15$&$-1.1$&$-1.3$&$0.0$&$-0.1$&$0.0$&$0.1$&$0.0$&$0.0$\\
$0$&$-1.3$&$-1.3$&$0.0$&$-0.1$&$0.0$&$0.1$&$0.0$&$0.0$\\
$-15$&$-1.4$&$-1.2$&$0.1$&$-0.1$&$0.0$&$0.1$&$0.0$&$0.0$\\
$-30$&$-1.6$&$-1.1$&$0.2$&$0.0$&$0.0$&$0.1$&$0.0$&$0.0$\\
$-45$&$-1.8$&$-1.0$&$0.3$&$0.1$&$0.0$&$0.0$&$0.0$&$0.0$\\
$-60$&$-2.1$&$-0.9$&$0.5$&$0.3$&$0.0$&$0.0$&$0.0$&$0.0$\\
$-75$&$-3.1$&$-0.4$&$0.9$&$1.3$&$0.3$&$0.3$&$0.0$&$0.0$\\
$-85$&$-6.8$&$1.5$&$1.0$&$9.3$&$8.2$&$3.0$&$0.0$&$0.0$\\
\hline
\end{tabular}
\end{center}
\end{table}

\color{black}

\section{The influence of radial velocity}\label{sec:rv_influence}

\subsection{Principle of the method}\label{sec:principle}

\color{black}

\begin{figure}[t]
\color{black}
  \centering
  \includegraphics[width = \columnwidth]{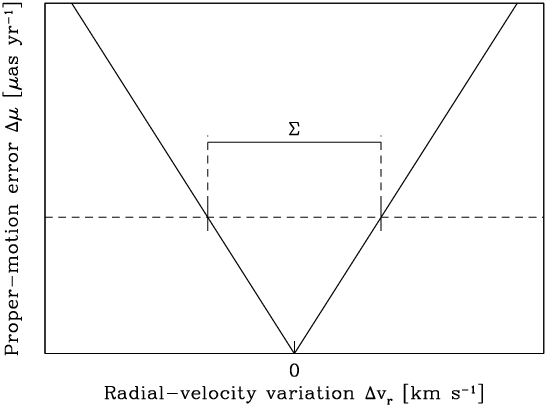}
  \caption{\color{black}Schematic diagram showing how to quantify the sensitivity of the proper motion to a change in (read: measurement error of) radial velocity. A change in the radial velocity $\Delta v_r$ introduced before the \color{black} backward solution \color{black} leads to a certain (HTPM) proper-motion error $\Delta\mu$. The linear dependence is commented on in Section~\ref{sec:sensitivity_derivation}. Since the magnitude of the proper-motion error does not depend on the sign but only on the magnitude of the radial-velocity variation, the sensitivity curve is symmetric with respect to the true radial velocity. The dashed horizontal line denotes the maximum perspective-acceleration-induced proper-motion error we are willing to accept in the HTPM proper motion. The distance $\Sigma$ between the intersection points of the dashed horizontal line and the solid sensitivity curves determines the tolerance on the radial-velocity error.}\label{fig:schematic}\color{black}
\end{figure}\color{black}

To quantify the influence of radial velocity on the HTPM proper motion for a star, we take the Hipparcos astrometric data (with epoch $1991.25$) and the literature radial velocity from the XHIP catalogue (Section~\ref{sec:XHIP} -- see Section \ref{sec:stars_without_radial_velocity} for stars without literature radial velocity) and use Equations (\ref{eq:forward1})--(\ref{eq:forward2}) to predict the star's celestial position in $2014.5$\footnote{To account for flexibility in the launch schedule of Gaia and to be on the safe side, we actually used a time interval of $25$ years, i.e., $2016.25$.}, i.e., the mean epoch of the first intermediate-release Gaia astrometry. We then use \color{black} backward solution\color{black}, i.e., apply Equations (\ref{eq:backward1})--(\ref{eq:backward2}), to recover the proper motion from the given Hipparcos and Gaia positions on the sky, assuming that the parallax and time interval are known. Clearly, if the radial velocity (radial proper motion) used in the \color{black} backward solution \color{black} is identical to the radial velocity used in the forward propagation, the derived (HTPM) proper motion is essentially identical to the input (XHIP) proper motion (Table~\ref{tab:error}). However, by varying the radial velocity used for the \color{black} backward solution \color{black} away from the input value, the sensitivity of the HTPM proper motion on radial velocity is readily established. This sensitivity does not depend on the sign but only on the magnitude of the radial-velocity variation. Figure~\ref{fig:schematic} schematically shows this idea. The abscissa shows the change in radial velocity, $\Delta v_r$, with respect to the input value used for the forward propagation. The ordinate shows $\Delta\mu$ (either $\Delta\mu_{\alpha}$ or $\Delta\mu_{\delta}$), i.e., the difference between the input value of the proper motion (either $\mu_{\alpha}$ or $\mu_{\delta}$) and the HTPM proper motion derived from the \color{black} backward solution\color{black}. The dashed horizontal line represents the maximum perspective-acceleration-induced proper-motion error that we are willing to accept. \color{black} If we denote the radial-velocity interval spanned by the intersections between the dashed, horizontal threshold line and the two solid sensitivity curves by $\Sigma$ (either $\Sigma_{\alpha}$ or $\Sigma_{\delta}$), the tolerance on the radial-velocity standard error is easily expressed as $-\frac{1}{2} \cdot \Sigma < \Delta v_r < \frac{1}{2} \cdot \Sigma$. The question now is: what is the probability that the error in radial velocity (i.e., true radial velocity minus catalogue value) is smaller than $-\frac{1}{2} \cdot \Sigma$ or larger than $\frac{1}{2} \cdot \Sigma$? Naturally, we want this probability to be smaller than a chosen threshold $1 - c$, where $c$ denotes the confidence level (for instance $c = 0.6827$ for a '$1$$\sigma$ result'):
\begin{equation}
P\left(-\frac{\Sigma}{2} < \Delta v_r < \frac{\Sigma}{2} \right) = \color{black}\tilde{\Phi}\left[{{\Sigma}\over{2 \cdot \sigma_{v_r}}}\right] \color{black} > c,
\label{eq:c_without_errors}
\end{equation}
\color{black} where we have assumed that the error distribution for $\Delta V_r$ is a normal distribution with zero mean and variance $\sigma_{v_r}^2$ \color{black} and where \color{black} $\tilde{\Phi}(x)$ \color{black} denotes the error function with argument \color{black}$x~ / \sqrt{2}$\color{black}:
\color{black}
\begin{equation}
\tilde{\Phi}(x) = {{\sqrt{2}}\over{\sqrt{\pi}}} \cdot \int_{0}^{x} \exp\left[-\frac{1}{2} \cdot t^2\right] {\rm d}t = {\rm erf}\left({{x}\over{\sqrt{2}}}\right).\label{eq:erf}
\end{equation}
\color{black}
From Equation~(\ref{eq:c_without_errors}), one can easily deduce:
\begin{equation}
\sigma_{v_r} < \color{black} {{\Sigma}\over{2 \cdot \tilde{\Phi}^{-1}(c)}},\color{black}
\label{eq:sigma_vr_without_errors}
\end{equation}
where \color{black} $\tilde{\Phi}^{-1}$ \color{black} denotes the inverse \color{black} of $\tilde{\Phi}$ \color{black} (e.g\color{black}.\color{black}, \color{black} $\tilde{\Phi}^{-1}[0.6827] = 1$\color{black}). For the 'special case' $c = 68.27\%$, Equation~(\ref{eq:sigma_vr_without_errors}) hence simplifies to $\sigma_{v_r} < \frac{1}{2} \cdot \Sigma$.\color{black}

\color{black}

\begin{figure}[t]
  \centering
  \includegraphics[width = \columnwidth]{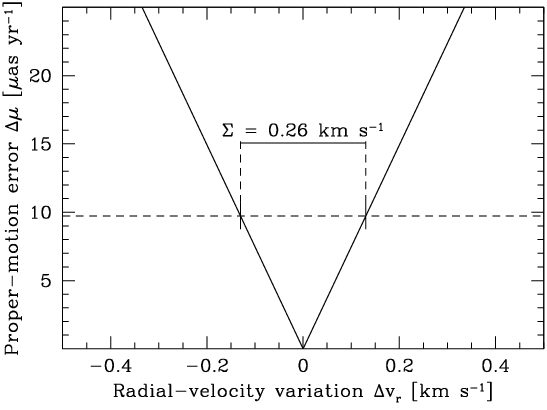}
  \color{black}
  \caption{\color{black}Sensitivity of the HTPM proper motion in right ascension to radial velocity for HIP$70890$ (Proxima Centauri). The sensitivity is linear and has a value \color{black} $C_\alpha = -74.59~\mu$as~yr$^{-1}$ per km~s$^{-1}$ \color{black} (Section~\ref{sec:sensitivity_derivation}). The dashed horizontal line indicates the maximum-tolerable perspective-acceleration-induced proper-motion error caused by an incorrect radial velocity. Since the expected HTPM standard error in right ascension is $97~\mu$as~yr$^{-1}$ for this star, we set this threshold to $97 / 10 = 9.7~\mu$as~yr$^{-1}$. This implies\color{black}, for a confidence level $c = 68.27\%$, \color{black} that the maximum-acceptable radial-velocity standard error $\sigma_{v_r}$ for this object is $\frac{1}{2} \cdot \Sigma = 0.13$~km~s$^{-1}$.}\label{fig:HIP70890}
  \color{black}
\end{figure}

\color{black}

\subsection{Example for a real star}\label{sec:HIP70890}

Figure~\ref{fig:HIP70890} is similar to the schematic Figure~\ref{fig:schematic} but shows data for a real star, HIP$70890$. This object, also known as Proxima Centauri (or $\alpha$~Cen~C), is a nearby \color{black} ($\varpi = 772.33 \pm 2.60$~mas), fast-moving ($\mu_\alpha = -3775.64 \pm 1.52$~mas~yr$^{-1}$ and $\mu_\delta = 768.16 \pm 1.82$~mas~yr$^{-1}$) \color{black} M6Ve flaring emission-line star which is known to have a significant perspective acceleration. HIP$70890$ is actually one of the 21 stars in the Hipparcos catalogue for which the perspective acceleration was taken into account \citep[Section 1.2.8]{1997ESASP1200.....P}. Figure~\ref{fig:HIP70890} was constructed by comparing the input proper motion used in the forward propagation with the proper motion resulting from the \color{black} backward solution \color{black} while using a progressively differing radial velocity in the \color{black} backward solution \color{black} from the (fixed) value used in the forward propagation. The actual radial-velocity variation probed in this figure is small, only $\pm 0.5$~km~s$^{-1}$.

HIP$70890$ is relatively faint ($H\!p = 10.7613$~mag) and the expected HTPM proper-motion standard error is $97~\mu$as~yr$^{-1}$ (see Section~\ref{sec:threshold} for details). If using a ten times lower threshold, i.e., $9.7~\mu$as~yr$^{-1}$, for the perspective-acceleration-induced proper-motion error caused by an incorrect radial velocity (see Section~\ref{sec:threshold} for details), we find that $\Sigma = 0.26$~km~s$^{-1}$. This implies\color{black}, for a confidence level $c = 68.27\%$, \color{black} that the radial velocity should \color{black} have been measured for this object with a standard error smaller than $\sigma_{v_r} < \color{black} {{\Sigma}\over{2 \cdot \tilde{\Phi}^{-1}(c = 0.6827)}} \color{black} = \frac{1}{2} \cdot \Sigma = 0.13$~km~s$^{-1}$. \color{black} The literature radial velocity for this star is $v_r = -22.40 \pm 0.50$~km~s$^{-1}$ (with quality grade 'B'; Section~\ref{sec:XHIP_radial_velocities}), which is not precise enough. New spectroscopic measurements are thus needed for this object to reduce the standard error by a factor $\sim$$4$.

The discussion above has implicitly focused on the right-ascension proper-motion component $\mu_{\alpha}$, and the associated $\Sigma_\alpha$, since the sensitivty of $\mu_{\delta}$ is a factor $\sim$$4$ less stringent for this star. It is generally sufficient to consider the most constraining case for a given star, i.e., either $\Sigma_{\alpha}$ or $\Sigma_{\delta}$. Therefore, we drop from here on the subscript $\alpha$ and $\delta$ on $\Sigma$, implicitly meaning that it either refers to $\Sigma_{\alpha}$ or $\Sigma_{\delta}$, depending on which one is  largest. Typically, this is the largest proper-motion component, i.e., \color{black} $|\mu_\alpha| > |\mu_\delta| \rightarrow \Sigma = \Sigma_\alpha$ and $|\mu_\alpha| < |\mu_\delta| \rightarrow \Sigma = \Sigma_\delta$\color{black}.

\begin{figure}[t]
  \centering
  \includegraphics[width = \columnwidth]{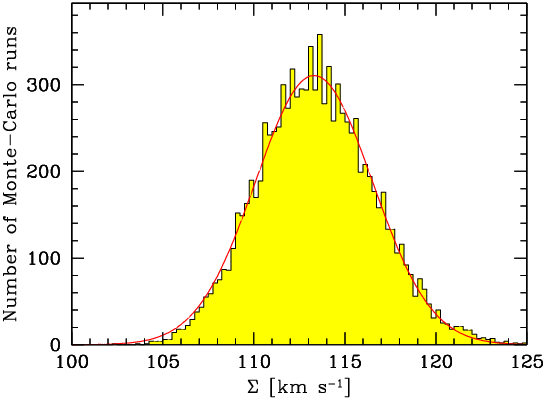}
  \caption{Histogram of the distribution of $\Sigma$ in the $N = 10,000$ Monte-Carlo simulations for star HIP$38$ ($\varpi = 23.64 \pm 0.66$~mas, so 3\% relative error). The smooth, red curve is a Gaussian fit of the histogram; it provides a good representation.}\label{fig:HIP38_gauss}
\end{figure}

\subsection{Derivation of the sensitivity}\label{sec:sensitivity_derivation}

\begin{figure}[t]
  \centering
  \includegraphics[width = \columnwidth]{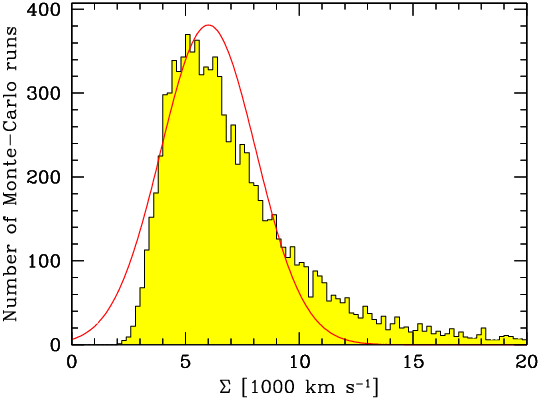}
  \caption{Histogram of the distribution of $\Sigma$ in the $N = 10,000$ Monte-Carlo simulations for star HIP$8$ ($\varpi = 4.98 \pm 1.85$~mas, so 37\% relative error). The smooth, red curve is a Gaussian fit of the histogram; it provides a poor representation and does not account for the tail in the distribution.}\label{fig:HIP8_tail}
\end{figure}

Figure~\ref{fig:HIP70890} shows that the sensitivity of proper motion to radial velocity is linear. This can be understood by substitution of Equations~(\ref{eq:backward1})--(\ref{eq:backward2}) in Equations~(\ref{eq:forward1})--(\ref{eq:forward2}), after replacing $v_r$, as used in the forward propagation, by $v_r + \Delta v_r$ in the \color{black} backward solution\color{black}:
\begin{eqnarray}
\phantom{0}
\Delta\mu_{\alpha} & = & \mu_{\alpha} - \frac{1}{t} \cdot \left[\Delta\alpha^* + \Delta\alpha^*\left(\mu_r+\frac{\Delta v_r}{r}\right)t-\tan\delta_0\Delta\alpha^*\Delta\delta + \right. \nonumber\\
& \phantom{=} & \left. + \frac{3\cos^2\delta_0-1}{6\cos^2\delta_0}(\Delta\alpha^*)^3-\tan\delta_0\Delta\alpha^*\Delta\delta\left(\mu_r+\frac{\Delta v_r}{r}\right)t\right]\nonumber\\
& = & \mu_{\alpha} - \frac{1}{t}\cdot \left[\mu_{\alpha}t +\Delta\alpha^*\frac{\Delta v_r}{r}t-\tan\delta_0\Delta\alpha^*\Delta\delta\frac{\Delta v_r}{r}t\right]\nonumber\\
& = & \underbrace{\left[\frac{\Delta\alpha^*}{r}-\tan\delta_0\Delta\alpha^*\Delta\delta\cdot \frac{1}{r}\right]}_{C_{\alpha}} \cdot \Delta v_r,\label{eq:Calpha}
\end{eqnarray}
\color{black}
which immediately shows $\Delta\mu_{\alpha} \propto \Delta v_r$. A similar analysis for the sensitivity coefficient $C_{\delta}$ yields:
\begin{equation}
\Delta\mu_{\delta} = \underbrace{\left[\frac{\Delta\delta}{r} + \frac{1}{2}\tan\delta_0(\Delta\alpha^*)^2\cdot \frac{1}{r}\right]}_{C_{\delta}} \cdot \Delta v_r.\label{eq:Cdelta}
\end{equation}
The coefficients $C_{\alpha}$ and $C_{\delta}$ quantify the proper-motion-estimation error caused by a biased knowledge of $v_r$ given measured displacements $\Delta\alpha^*$ and $\Delta\delta$. They can hence more formally be defined as the partial derivatives of $\mu_\alpha$ and $\mu_\delta$ from Equations (\ref{eq:backward1}) and (\ref{eq:backward2}) with respect to $v_r$ with $\Delta\alpha^*$ and $\Delta\delta$ kept constant:
\begin{eqnarray}
C_\alpha \equiv {{\partial \mu_\alpha}\over{\partial v_r}} = \frac{1}{r}\cdot{{\partial \mu_\alpha}\over{\mu_r}} &=& \frac{1}{r}\cdot\left[\Delta\alpha^*-\tan\delta_0\Delta\alpha^*\Delta\delta\right]\\
C_\delta \equiv {{\partial \mu_\delta}\over{\partial v_r}} = \frac{1}{r}\cdot{{\partial \mu_\delta}\over{\mu_r}} &=& \frac{1}{r}\cdot\left[\Delta\delta+\frac{1}{2}\tan\delta_0(\Delta\alpha^*)^2\right].
\end{eqnarray}
Clearly, our relations confirm the well-known, classical result \citep[e.g.,][Equation~4]{1999A&A...348.1040D} that perspective-acceleration-induced proper-motion errors are proportional to the product of the time interval, the parallax $\varpi \propto r^{-1}$, the proper-motion components $\mu_{\alpha, \delta}$, and the radial-velocity error $\Delta v_r$: $\Delta\mu_{\alpha,\delta} = C_{\alpha,\delta} \cdot \Delta v_r = r^{-1} \cdot \mu_{\alpha,\delta} t \cdot \Delta v_r$. Equations~(\ref{eq:Calpha})--(\ref{eq:Cdelta}) show that the perspective-acceleration-induced proper-motion error caused by a radial-velocity error does not depend on the radial velocity $v_r$ itself but only on the error $\Delta v_r$. \color{black} This may look counter-intuitive at first sight, since the proper motion itself {\it is} sensitive to the precise value of the radial velocity. The error in the proper motion, however, is sensitive only to the radial-velocity error. In other words, the slopes of the V-shaped wedge in Figure~\ref{fig:HIP70890} do not depend on the absolute but only on the relative labeling of the abscissa.

\subsection{Taking measurement errors into account}\label{sec:measurement_errors}

\color{black} So far, we ignored the measurement errors of the astrometric parameters $\alpha$, $\delta$, $\varpi$, $\mu_{\alpha}$, and $\mu_{\delta}$. \color{black} A natural way to take these errors into account is by Monte-Carlo simulations: rather than deriving $\Sigma$ once, namely based on the astrometric parameters contained in the XHIP catalogue, we calculate $\Sigma$ a large number of times (typically $N = 10,000$), where in each run we do not use the catalogue astrometry but randomly distorted values drawn from normal distributions centred on the measured astrometry and with standard deviations equal to the standard errors of the astrometric parameters (denoted \color{black} $N({\rm mean}, {\rm variance})$\color{black}). We also randomly draw the radial velocity in each run from the normal probability distribution \color{black}$N(v_r, \sigma_{v_r}^2)$\color{black}.

The Monte-Carlo simulations yield $N = 10,000$ values for $\Sigma$; the interpretation of this distribution will be addressed in Section~\ref{sec:evaluation_criteria}. Two representative examples of the distribution of $\Sigma$ are shown in Figures~\ref{fig:HIP38_gauss} and \ref{fig:HIP8_tail}. The first distribution (Figure~\ref{fig:HIP38_gauss}, representing HIP$38$) is for a nearby star with a well-determined parallax: $\varpi = 23.64 \pm 0.66$~mas, i.e., 3\% relative parallax error. The smooth, red curve in Figure~\ref{fig:HIP38_gauss} is a Gaussian fit of the histogram; it provides a good representation of the data. The second distribution (Figure~\ref{fig:HIP8_tail}, representing HIP$8$) is for a distant star with a less well-determined parallax: $\varpi = 4.98 \pm 1.85$~mas, i.e., 37\% relative parallax error. This results in an asymmetric distribution of $\Sigma$ with a tail towards large $\Sigma$ values. \color{black} This is easily explained since we essentially have $\Sigma \propto \Delta v_r \propto C_{\alpha,\delta}^{-1} \propto \varpi^{-1}$, meaning that the distribution of $\Sigma$ reflects the probability distribution function of $\varpi^{-1} \propto r$. The latter is well-known \citep[e.g.,][]{1998A&A...340L..35K,1999ASPC..167...13A} for its extended tail towards large distances and its (light) contraction for small distances. \color{black} The smooth, red curve in Figure~\ref{fig:HIP8_tail} is a Gaussian fit of the histogram; it provides an inadequate representation of the data. We will come back to this in Section~\ref{sec:robust_criterion}.

\color{black} To avoid dealing with (a significant number of) negative parallaxes in the Monte-Carlo simulations, we decided to ignore $11,171$ entries with insignificant parallax measurements in the XHIP catalogue; these include $3,920$ entries with $\varpi \leq 0$ and $7,251$ entries with $0 < \varpi / \sigma_\varpi \leq 1$ \citep[recall that negative parallaxes are a natural outcome of the Hipparcos astrometric data reduction, e.g.,][]{1995A&A...304...52A}. This choice does not influence the main conclusions of this paper: perspective-acceleration-induced HTPM proper-motion errors are significant only for nearby stars whereas negative and low-significance parallax measurements generally indicate large distances.
\color{black}

\section{Evaluation criteria}\label{sec:evaluation_criteria}

From the Monte-Carlo distribution of $\Sigma$ (Section~\ref{sec:measurement_errors}), we want to extract information to decide whether the radial velocity available in the literature is sufficiently precise or not. For this, we develop two evaluation criteria.

\subsection{The Gaussian evaluation criterion}\label{sec:gaussian_criterion}

The first criterion, which we refer to as the Gaussian criterion, is based on Gaussian interpretations of probability distributions. It can be applied to all stars but, since we have seen that distant stars do not have a Gaussian $\Sigma$ distribution, but rather a distribution with tails towards large $\Sigma$ values (Figure~\ref{fig:HIP8_tail}), the Gaussian criterion is unbiased, and hence useful, only for nearby stars. \color{black} For these stars, the Monte-Carlo distribution of $\Sigma$ is well described by a Gaussian function with mean $\mu_\Sigma$ and standard deviation $\sigma_\Sigma$ (Figure~\ref{fig:HIP38_gauss}). The equations derived in Section~\ref{sec:principle} by ignoring astrometric errors are easily generalised by recognising that both the radial-velocity error and the distribution of $\Sigma$ have Gaussian distributions, with standard deviations $\sigma_{v_r}$ and $\sigma_\Sigma$, respectively:
\begin{equation}
P\left(-\frac{\Sigma}{2} < \Delta v_r < \frac{\Sigma}{2}\right) = \color{black} \tilde{\Phi}\left[{{\Sigma}\over{2 \cdot \sigma_{v_r}}}\right] \color{black} > c
\end{equation}
generalises to:
\begin{equation}
\int\limits_{\Sigma=-\infty}^{\Sigma=\infty} {{{\rm d}\Sigma}\over{\sqrt{2\pi\sigma_\Sigma^2}}} \cdot \exp\left[-\frac{1}{2}\cdot\left({{\Sigma-\mu_\Sigma}\over{\sigma_\Sigma}}\right)^2\right] \cdot \color{black} \tilde{\Phi}\left[{{\Sigma}\over{2 \cdot \sigma_{v_r}}}\right] \color{black} > c
\label{eq:c_gauss_original}
\end{equation}
Using Equation 8.259.1 from \cite{2007gr...book.....G}, Equation~(\ref{eq:c_gauss_original}) simplifies to:
\begin{equation}
\color{black} \tilde{\Phi}\left[{{\mu_\Sigma}\over{\sqrt{4\cdot\sigma_{v_r}^2 + \sigma_\Sigma^2}}}\right] \color{black} > c,
\label{eq:c_gauss}
\end{equation}
which correctly reduces to Equation~(\ref{eq:c_without_errors}) for the limiting, 'error-free' case $\mu_\Sigma \rightarrow \Sigma$ and $\sigma_\Sigma \rightarrow 0$ in which the Gaussian distribution of $\Sigma$ collapses into a delta function at $\mu_\Sigma = \Sigma$, i.e., $\delta(\Sigma)$.

For the example of HIP$70890$ discussed in Section~\ref{sec:HIP70890}, we find $\mu_\Sigma = 0.26$~km~s$^{-1}$ and $\sigma_\Sigma = 0.000,89$~km~s$^{-1}$ so that, with $\sigma_{v_r} = 0.50$~km~s$^{-1}$, Equation~(\ref{eq:c_gauss}) returns $c = 20.58\%$.

It is trivial, after re-arranging Equation~(\ref{eq:c_gauss}) to:
\begin{equation}
\sigma_{v_r} < \color{black} \sqrt{{{\mu_\Sigma^2}\over{4\cdot[\tilde{\Phi}^{-1}(c)]^2}}-{{\sigma_\Sigma^2}\over{4}}}\color{black},
\label{eq:sigma_vr_gauss}
\end{equation}
to compute the required standard error of the radial velocity to comply with a certain confidence level $c$. For instance, if we require a $c = 99.73\%$ confidence level (for a '$3$$\sigma$ result'), the radial-velocity standard error of HIP$70890$ has to be $0.04$~km~s$^{-1}$. We finally note that Equation~(\ref{eq:sigma_vr_gauss}) correctly reduces to Equation~(\ref{eq:sigma_vr_without_errors}) for the limiting, 'error-free' case $\mu_\Sigma \rightarrow \Sigma$ and $\sigma_\Sigma \rightarrow 0$.

A limitation of Equation~(\ref{eq:sigma_vr_gauss}) is that the argument of the square root has to be non-negative. This is physically easy to understand when realising that, in the Gaussian approximation, one has $\sigma_\Sigma \approx \mu_\Sigma \cdot (\sigma_\varpi / \varpi)$ (see Section~4.4), so that:
\begin{eqnarray}
\color{black}
{{\mu_\Sigma^2}\over{4\cdot[\tilde{\Phi}^{-1}(c)]^2}}-{{\sigma_\Sigma^2}\over{4}} \geq 0 \rightarrow
\tilde{\Phi}^{-1}(c) \leq {{\varpi}\over{\sigma_\varpi}} \rightarrow c \leq \tilde{\Phi}\left[{{\varpi}\over{\sigma_\varpi}}\right]\color{black}.
\end{eqnarray}
So, for instance, if a certain star has $\varpi / \sigma_\varpi = 2$ (a '$2$$\sigma$ parallax'), the Gaussian methodology will only allow to derive the radial-velocity standard error $\sigma_{v_r}$ required to meet a confidence level $c = 95.45\%$ or lower.
\color{black}

\subsection{The robust criterion}\label{sec:robust_criterion}

Since the Monte-Carlo distribution of $\Sigma$ values is not Gaussianly distributed for distant stars (Figure~\ref{fig:HIP8_tail}), the Gaussian criterion returns incorrect estimates; in fact, the estimates are not just incorrect but also biased since the Gaussian criterion systematically underestimates the mean value of $\Sigma$ (i.e., \color{black}$\mu_\Sigma$\color{black}) and hence systematically provides too conservative \color{black} (small) \color{black} estimates for $\sigma_{v_r}$ through Equation~(\ref{eq:sigma_vr_gauss}). Rather than fitting a Gaussian function, we need a more robust estimator of the location and width of the $\Sigma$ distribution than the Gaussian mean and standard deviation. This estimator is contained in the data itself and provides, what we call, the robust criterion.

\begin{figure}[t]
  \centering
  \includegraphics[width = \columnwidth]{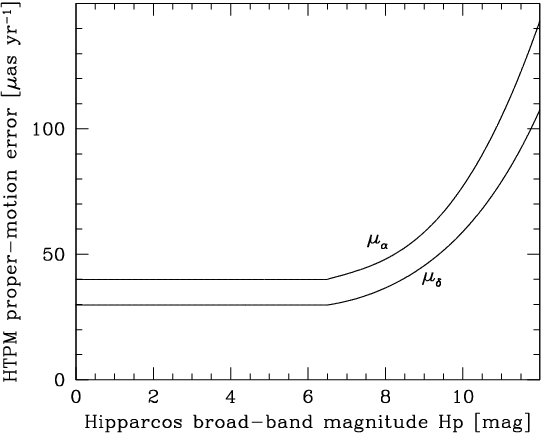}
  \caption{Predicted HTPM proper-motion error as function of the $H\!p$ Hipparcos broad-band magnitude following Equations~(\ref{eq:threshold_alpha})--(\ref{eq:threshold_delta}). We require perspective-acceleration-induced proper-motion errors to be an order of magnitude smaller (${\rm factor\ of\ safety} = {\rm FoS} = 10$; see Section ~\ref{sec:threshold_discussion}).}\label{fig:HTPM_precision}
\end{figure}

\color{black}
Let us denote the individual values of $\Sigma$ derived from the $N = 10,000$ Monte-Carlo simulations by $\Sigma_i$, with $i = 1, \ldots, N$. Equation~(\ref{eq:c_gauss_original}) is then readily generalised for arbitrary distributions of $\Sigma$ to:
\begin{equation}
\sum\limits_{i = 1}^{N} {{1}\over{N}} \cdot P\left(-\frac{\Sigma_i}{2} < \Delta v_r < \frac{\Sigma_i}{2}\right) = \sum\limits_{i = 1}^{N} {{1}\over{N}} \cdot \color{black} \tilde{\Phi}\left[{{\Sigma_i}\over{2 \cdot \sigma_{v_r}}}\right] \color{black} > c.
\label{eq:original_robust}
\end{equation}
The inverse relation generalising Equation~(\ref{eq:sigma_vr_gauss}) by expressing $\sigma_{v_r}$ as function of $c$, required to determine the precision of the radial velocity required to comply with a certain confidence level $c$, is not analytical; we hence solve for it numerically.
\color{black}

The robust criterion generalises the Gaussian criterion. Both criteria return the same results for nearby stars which have a symmetric Gaussian distribution of $\Sigma$ values. In general, therefore, the robust criterion is the prefered criterion for all stars, regardless of their distance.

\section{Application of the evaluation criteria}\label{sec:application}

\subsection{Target proper-motion-error threshold}\label{sec:threshold}

Before we can apply the Gaussian and robust criteria, we have to decide on a target proper-motion-error threshold for each star (i.e., the location of the dashed horizontal lines in Figures~\ref{fig:schematic} and \ref{fig:HIP70890}). We adopt as a general rule that the maximum perspective-acceleration-induced HTPM proper-motion error caused by radial-velocity errors shall be an order of magnitude smaller than the predicted standard error of the HTPM proper motion itself (Section~\ref{sec:threshold_discussion} discusses this choice in more detail). The latter quantity has been studied by \citet{FM-040} and can be parametrised as:
\begin{eqnarray}
\sigma_{\mu_{\alpha}}\ [\mu{\rm as~yr}^{-1}] &=& -227.8 + 122.1 \cdot H - 20.39 \cdot H^2\nonumber\\
&& + 1.407 \cdot H^3 - 0.02841 \cdot H^4\label{eq:threshold_alpha}\\
\sigma_{\mu_{\delta}}\ [\mu{\rm as~yr}^{-1}] &=& \phantom{-}127.2 - \phantom{0}47.0 \cdot H + \phantom{0}8.30 \cdot H^2\nonumber\\
&& - 0.686 \cdot H^3 + 0.02581 \cdot H^4\label{eq:threshold_delta},
\end{eqnarray}
where $H = \max\{6.5, H\!p\ {\rm [mag]}\}$ with $H\!p$ the Hipparcos broad-band magnitude. These relations are shown in Figure~\ref{fig:HTPM_precision}. The predicted HTPM standard errors include residual errors caused by the correction for the parallactic effect in the Gaia data (see footnote~1), the expected number and temporal distribution of the Gaia field-of-view transits for the Gaia nominal sky scanning law, and the expected location-estimation precision ('centroiding error') of Gaia's CCD-level data.

\subsection{Stars without literature radial velocities}\label{sec:stars_without_radial_velocity}

\color{black}

\begin{figure}[t]
  \includegraphics[width = \columnwidth]{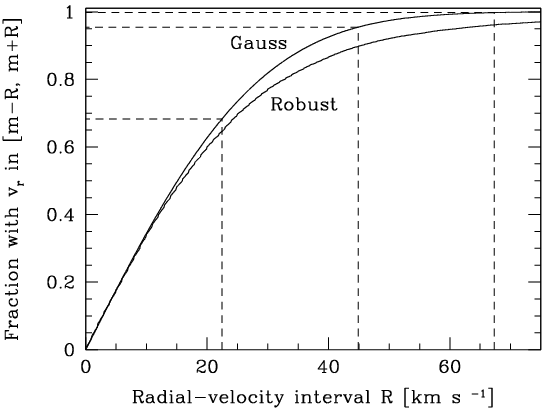}
  \color{black}
  \caption{\color{black}The fraction of stars with XHIP literature radial velocities which are contained in the radial-velocity interval $[m-R, m+R]$ as function of $R$, with $m = -2.21$~km~s$^{-1}$ the mean $v_r$ for the Gaussian criterion and $m = -2.00$~km~s$^{-1}$ the median $v_r$ for the robust criterion (Sections~\ref{sec:XHIP_radial_velocities} and \ref{sec:stars_without_radial_velocity}). For the Gaussian criterion, we represent the histogram of literature radial velocities by a Gauss with standard deviation $\sigma = 22.44$~km~s$^{-1}$ (Figure~\ref{fig:all_vr}). The dashed lines represent the classical limits $1\sigma = 68.27\%$, $2\sigma = 95.45\%$, and $3\sigma = 99.73\%$. The fraction of stars with the robust criterion builds up more slowly as a result of the non-Gaussian broad wings as well as outliers representing halo and runaway stars. Since the Gaussian criterion ignores these features, it returns biased results for stars without literature radial velocity (see Section~\ref{sec:application_results}).\color{black}}\label{fig:vrad_errors}
\end{figure}

\color{black}

\begin{figure}[t]
  \includegraphics[width = \columnwidth]{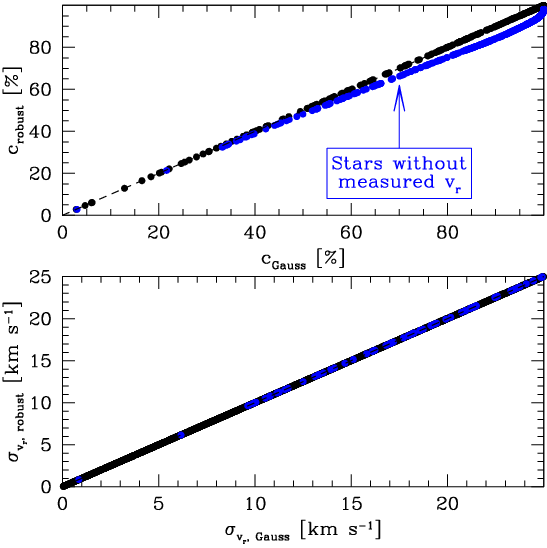}
  \color{black}
  \caption{\color{black}Comparison of the robust and Gaussian criteria for all stars with $\sigma_\varpi / \varpi$ better than $5$\%. The top panel compares the \color{black} confidence levels \color{black} while the bottom panel compares the radial-velocity standard errors $\sigma_{v_r}$ required to reach a \color{black} confidence level $c = 68.27\%$\color{black}. The top panel shows two branches of data points: the linear, one-to-one branch corresponds to stars with a measured radial velocity, whereas the lower, curved branch corresponds to stars without measured radial velocity in the XHIP catalogue. As explained in Section~\ref{sec:stars_without_radial_velocity}, the latter objects suffer from a bias in the Gaussian \color{black} confidence level $c_{\rm Gauss}$\color{black}.}\label{fig:comparison}
\end{figure}

\color{black}

As already discussed in Section~\ref{sec:XHIP_radial_velocities}, the majority of stars in the XHIP catalogue do not have a literature radial velocity. These $71,563$ objects are treated as stars with radial velocity, with three exceptions:
\begin{enumerate}
\item For each of the $N = 10,000$ Monte-Carlo simulations, the radial velocity is randomly taken from the full list of $46,392$ radial velocites contained in the XHIP catalogue (Figure~\ref{fig:all_vr}). In practice, this choice does not influence our results since we are not sensitive to the absolute value of the radial velocity (Section~\ref{sec:sensitivity_derivation}). But at least in principle this choice means that there is a finite probability to assign a halo-star-like, i.e., large radial velocity in one (or more) of the Monte-Carlo runs. \color{black} Regarding the HTPM Project, the best choice for stars without known radial velocity is to use $v_r = -2.00$~km~s$^{-1}$, which is the median value of the distribution (recall that the mean equals $-2.21$~km~s$^{-1}$). \color{black}
\item \color{black} For the Gaussian evaluation criterion (Section~\ref{sec:gaussian_criterion}), we use $\sigma_{v_r} = 22.44$~km~s$^{-1}$ from the Gaussian fit to the distribution of all literature radial velocities. \color{black} We also follow this recipe for the Gaussian criterion for the $1,753$ entries which do have a radial velocity but which do not have an associated standard error in the XHIP catalogue (this concerns $23$ entries with quality grade 'C' and $1,730$ entries with quality grade 'D'). Clearly, this approach ignores the broad wings of the distribution visible in Figure~\ref{fig:all_vr} as well as a small but finite number of halo stars and runaway stars with heliocentric radial velocities up to plus-or-minus several hundred km~s$^{-1}$ (see Section~\ref{sec:XHIP_radial_velocities}). As a result, the Gaussian criterion systematically returns an overly optimistic (i.e., too large) \color{black} confidence level $c_{\rm Gauss}$ \color{black} for stars without literature radial velocity \color{black} (Figures~\ref{fig:vrad_errors} and \ref{fig:comparison})\color{black}. This bias comes in addition to the bias for distant stars for which the Gaussian criterion returns too conservative \color{black} (small) \color{black} estimates for $\sigma_{v_r}$ (Section~\ref{sec:robust_criterion}).
\item For the robust evaluation criterion (Section~\ref{sec:robust_criterion}), we do not make {\it a priori} assumptions except that the overall radial-velocity distribution of stars without known radial velocity is the same as the distribution of stars with literature radial velocity, including broad wings and 'outliers'. \color{black} We thus use Equation~(\ref{eq:original_robust}) in the form:
\begin{equation}
\sum\limits_{i = 1}^{N} {{1}\over{N}} \cdot P\left(-\frac{\Sigma_i}{2} < \Delta v_r < \frac{\Sigma_i}{2}\right) > c,
\end{equation}
where the probability $P$ that $\Delta v_r$ is contained in the interval $[-\frac{1}{2} \cdot \Sigma_i, \frac{1}{2} \cdot \Sigma_i]$ is calculated as the fraction of all stars with literature radial velocities in the XHIP catalogue which has $v_r \in [-\frac{1}{2} \cdot \Sigma_i + v_{r, {\rm median}}, \frac{1}{2} \cdot \Sigma_i + v_{r,{\rm median}}]$ (recall that the median radial velocity equals $-2.00$~km~s$^{-1}$). \color{black} We thus cater for the broad wings of the observed radial-velocity distribution \color{black} (Figures~\ref{fig:all_vr} and \ref{fig:vrad_errors}) \color{black} as well as for the probability that the object is a (fast-moving) halo star, avoiding the bias in the Gaussian criterion discussed in the previous bullet.
\end{enumerate}

\color{black}

\subsection{Results of the application}\label{sec:application_results}

\color{black}

We applied the Gaussian and robust criteria, as described in Sections~\ref{sec:gaussian_criterion} and \ref{sec:robust_criterion}, respectively, to the XHIP catalogue. The results are presented in Table~\ref{tab:datafile} (Appendix~\ref{sec:datafile}), which is available electronically only. The run time for $N = 10,000$ Monte-Carlo simulations is typically \color{black} $\sim$$0.7$~s \color{black} per star and processing the full set of \color{black} $117,955 - 11,171 = 106,784$ XHIP entries with significant parallaxes (Section~\ref{sec:measurement_errors}) hence takes about one day.\color{black}

\begin{figure}[t]
  \includegraphics[width = \columnwidth]{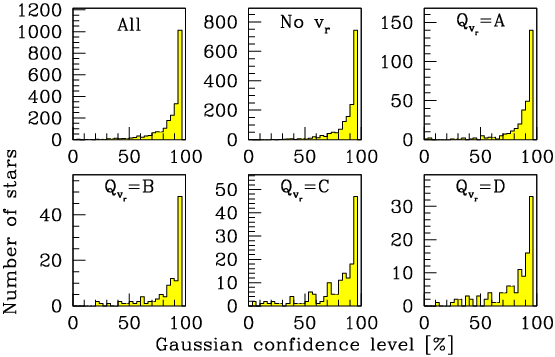}
  \color{black}
  \caption{\color{black}Histograms of the Gaussian \color{black} confidence level $c$ \color{black} from Section~\ref{sec:gaussian_criterion} for all stars combined, stars without measured radial velocity, and stars with measured radial velocity as function of radial-velocity quality grade $Q_{v_r}$ (see Section~\ref{sec:XHIP_radial_velocities} and Appendix~\ref{sec:datafile}). \color{black} The vast majority of objects have $c > 95.45$\% (see Table~\ref{tab:confidence_level_distribution_gaussian_criterion}); they have been omitted from the histograms to improve their legibility.\color{black}}\label{fig:confidence_level_histograms_gaussian_criterion}
  \color{black}
\end{figure}

\begin{figure}[t]
  \centering
  \color{black}
  \includegraphics[width = \columnwidth]{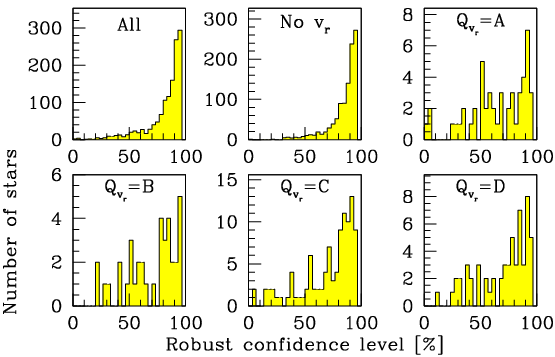}
  \caption{\color{black}As Figure~\ref{fig:confidence_level_histograms_gaussian_criterion}, but for the robust criterion from Section~\ref{sec:robust_criterion}.\color{black}}\label{fig:confidence_level_histograms_robust_criterion}
  \color{black}
\end{figure}

Figures~\ref{fig:confidence_level_histograms_gaussian_criterion}--\ref{fig:confidence_level_histograms_robust_criterion} and Tables~\ref{tab:confidence_level_distribution_gaussian_criterion}--\ref{tab:confidence_level_distribution_robust_criterion} show the results for the \color{black} confidence levels \color{black} of the literature radial velocities contained in the XHIP catalogue. We find, not surprisingly since perspective-acceleration-induced proper-motion errors are relevant only for nearby, fast-moving stars -- which are relatively rare -- that the majority of stars have \color{black} confidence \color{black} levels exceeding \color{black} $c = 99.73\%$\color{black}. This indicates that, at the $c = 99.73\%$ confidence level, the available radial velocity is sufficiently precise or, for stars without literature radial velocity, that the absence of a literature radial velocity, and hence the assumption $v_r = -2.00$~km~s$^{-1}$ for the robust criterion or $v_r = -2.21$~km~s$^{-1}$ for the Gaussian criterion, is acceptable. This holds for more than \color{black} $100,000$ \color{black} stars using the robust criterion (Table~\ref{tab:confidence_level_distribution_robust_criterion}) and \color{black} more than $85,000$ \color{black} stars using the Gaussian criterion (Table~\ref{tab:confidence_level_distribution_gaussian_criterion}). The large difference between the two criteria does not come unexpectedly:
\begin{enumerate}
\item We already argued in Section~\ref{sec:robust_criterion} that the Gaussian criterion is biased for distant stars, with distant meaning that the parallax probability distribution has an associated asymmetric distance probability distribution. For such stars, the Gaussian criterion systematically underestimates the mean value of $\Sigma$ and hence returns too conservative (small) values for $\sigma_{v_r}$ for a given value of \color{black} the confidence level $c$ \color{black} and too pessimistic (small) values of \color{black} $c$ \color{black} for a given value of $\sigma_{v_r}$;
\item We already argued in Section~\ref{sec:stars_without_radial_velocity} that the Gaussian criterion is biased for stars without literature radial velocity. For such stars, the Gaussian criterion systematically returns too optimistic (large) values of \color{black} the confidence level $c$ \color{black} since it ignores the broad wings of the observed distribution of radial velocities (\color{black}Figures~\ref{fig:all_vr}, \ref{fig:vrad_errors}, and \ref{fig:comparison}\color{black}) and also ignores the probability that the object is actually a halo (or runaway) star.
\end{enumerate}
The robust criterion does not suffer from these biases and hence, being more reliable, is prefered for all stars. The Gaussian criterion, nonetheless, provides a useful and also easily interpretable reference test case and we hence decided to retain it. Figure~\ref{fig:comparison} shows that, for nearby stars with literature radial velocities, the Gaussian and robust criteria return equivalent results.

For a small but non-negligible number of stars, Table~\ref{tab:confidence_level_distribution_robust_criterion} indicates insatisfactory results: \color{black} $206$ \color{black} stars have a \color{black} confidence level $c < 68.27\%$\color{black}: \color{black} $97$ \color{black} of these do have a literature radial velocity in the XHIP catalogue but one which is insufficiently precise. The remaining \color{black} $109$ \color{black} stars do not have a spectroscopically measured radial velocity (at least not one contained in the XHIP catalogue). New spectroscopy is hence required for these stars to guarantee a confidence level of at least $68.27\%$. For increased confidence levels, the numbers obviously increase: if requiring a \color{black} $c = 99.73\%$ \color{black} confidence level for all objects, for instance, the number of 'problem stars' increases to \color{black} $6,562$\color{black}, split into \color{black} $382$ \color{black} with insufficiently-precise known radial velocity and \color{black} $6,180$ \color{black} without known radial velocity. We conclude that, depending on the confidence level one wants to achieve, hundreds to thousands of stars need to be spectroscopically re-measured.

\color{black}

\begin{table}[t]
\caption{Number of stars as function of \color{black} confidence level $c$ \color{black} established using the Gaussian criterion from Section~\ref{sec:gaussian_criterion}. The number in brackets in column~2 indicates the number of bright stars ($G < 5.7$~mag or $H\!p < 6.0$~mag), i.e., those stars not detectable with Gaia and hence not contained in the HTPM catalogue (Section~\ref{sec:HTPM_bright_limit}). \color{black} Stars with insignificant parallax measurements have not been processed and the total number of entries hence equals $117,955 - 11,171 = 106,784$ (see Section~\ref{sec:measurement_errors}).\color{black}}\label{tab:confidence_level_distribution_gaussian_criterion}
\begin{center}
\color{black}
\begin{tabular}[h]{crrr}
\hline\hline
Conf.\ level $c$ [\%] & \# stars (\# bright) & $v_r \in$~XHIP & $v_r \not\in$~XHIP\\
\hline
$[0-68.27)$     & $    225~(\phantom{0,0}12)$& $   108$& $   117$\\
$[68.27-95.45)$ & $  2,087~(\phantom{0,0}27)$& $   598$& $ 1,489$\\
$[95.45-99.73)$ & $ 18,815~(\phantom{0,0}88)$& $ 4,104$& $14,711$\\
$[99.73-100]$   & $ 85,657~          (4,341)$& $38,700$& $46,957$\\
\hline
Total           & $106,784~          (4,468)$& $43,510$& $63,274$\\
\hline
\end{tabular}
\color{black}
\end{center}
\color{black}
\end{table}

\color{black}

\begin{table}[t]
\caption{Number of stars as function of \color{black} confidence level $c$ \color{black} established using the robust criterion from Section~\ref{sec:robust_criterion}. The number in brackets in column~2 indicates the number of bright stars ($G < 5.7$~mag or $H\!p < 6.0$~mag), i.e., those stars not detectable with Gaia and hence not contained in the HTPM catalogue (Section~\ref{sec:HTPM_bright_limit}). \color{black} Stars with insignificant parallax measurements have not been processed and the total number of entries hence equals $117,955 - 11,171 = 106,784$ (see Section~\ref{sec:measurement_errors}).\color{black}}\label{tab:confidence_level_distribution_robust_criterion}
\begin{center}
\color{black}
\begin{tabular}[h]{crrr}
\hline\hline
Conf.\ level $c$ [\%] & \# stars (\# bright) & $v_r \in$~XHIP & $v_r \not\in$~XHIP\\
\hline
$[0-68.27)$     & $    206~(\phantom{0,0}12)$& $    97$& $   109$\\
$[68.27-95.45)$ & $  1,112~(\phantom{0,0}15)$& $   150$& $   962$\\
$[95.45-99.73)$ & $  5,244~(\phantom{0,0}10)$& $   135$& $ 5,109$\\
$[99.73-100]$   & $100,222~          (4,431)$& $43,128$& $57,094$\\
\hline
Total           & $106,784~          (4,468)$& $43,510$& $63,274$\\
\hline
\end{tabular}
\color{black}
\end{center}
\color{black}
\end{table}

\color{black}

\color{black}

\color{black}

\begin{figure}[t]
  \centering
  \includegraphics[width = \columnwidth]{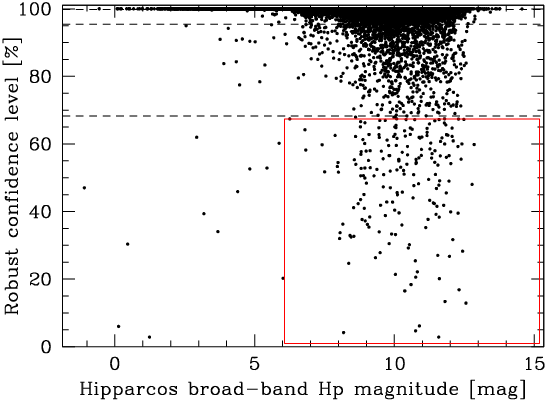}
  \color{black}
  \caption{\color{black}Robust \color{black} confidence level $c$ \color{black} versus Hipparcos broad-band $H\!p$ magnitude. The red box in the bottom-right corner denotes the approximate area for high-priority follow-up spectroscopy: stars with \color{black} confidence level $c < 68.27\%$ \color{black} and $H\!p > 6$~mag. The latter restriction roughly reflects Gaia's -- and hence HTPM's -- bright limit (Section~\ref{sec:HTPM_bright_limit}).\color{black}}\label{fig:mag}
  \color{black}
\end{figure}

\begin{figure}[t]
  \centering
  \includegraphics[width = \columnwidth]{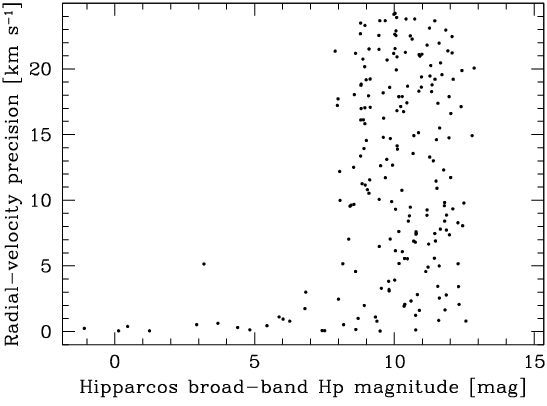}
  \color{black}
  \caption{\color{black}Radial-velocity precision (standard error) for stars with robust \color{black} confidence level $c < 68.27$\% \color{black} required to upgrade \color{black} their confidence level to $c = 68.27\%$\color{black}.\color{black}}\label{fig:acc_01}
  \color{black}
\end{figure}

Figure \ref{fig:mag} shows the robust \color{black} confidence level \color{black} versus $H\!p$ magnitude. One can see that the typical star which needs a high-priority spectroscopic measurement (i.e., \color{black} $c < 68.27\%$\color{black}) has $H\!p$ in the range $8$--$12$~mag. Figure~\ref{fig:acc_01} shows the radial-velocity precision required to reach \color{black} $c = 68.27\%$ \color{black} (computed with the robust criterion) versus magnitude. Precisions vary drastically, from very stringent values well below $1$~km~s$^{-1}$ to very loose values, up to several tens of km~s$^{-1}$.

\color{black}

\color{black}

\section{Discussion}\label{sec:discussion}

\subsection{HTPM bright limit}\label{sec:HTPM_bright_limit}

The HTPM catalogue will contain the intersection of the Hipparcos and Gaia catalogues. Whereas the Hipparcos catalogue \citep{1997yCat.1239....0E}, which contains $117,955$ entries with astrometry (and $118,218$ entries in total), is complete to at least $V = 7.3$~mag \citep[][Section~1.1]{1997ESASP1200.....P}, the Gaia catalogue will be incomplete at the bright end. Gaia's bright limit is $G = 5.7$~mag \citep{2012Ap&SS.tmp...68D}, where $G$ is the white-light, broad-band Gaia magnitude, which is linked to the Hipparcos $H\!p$, the Cousins $I$, and the Johnson $V$ magnitudes through \citep{2010A&A...523A..48J}:
\begin{eqnarray}
G - V &=& -0.0447 - 0.1634 \cdot (H\!p - I) + \nonumber\\
      & & +0.0331 \cdot (H\!p - I)^2 - 0.0371 \cdot (H\!p - I)^3.
\end{eqnarray}
For stars in the Hipparcos catalogue, the colour $G - H\!p$ ranges between $-0.5$ and $0.0$~mag, with a mean value of $-0.3$~mag. This means that $G = 5.7$~mag corresponds roughly to $H\!p \approx 6.0$~mag. In practice, therefore, we are not concerned with the brightest $\sim$$4,509$ stars in the sky. The expected number of HTPM-catalogue entries is therefore $\sim$$117,955 - 4,509 \approx 113,500$. \color{black} The number of entries with significant parallax measurements equals $117,955 - 11,171 = 106,784$, of which $106,784 - 4,468 = 102,316$ have $G > 5.7$~mag.\color{black}

\subsection{Proper-motion-error threshold}\label{sec:threshold_discussion}

\begin{figure}[t]
  \centering
  \includegraphics[width = \columnwidth]{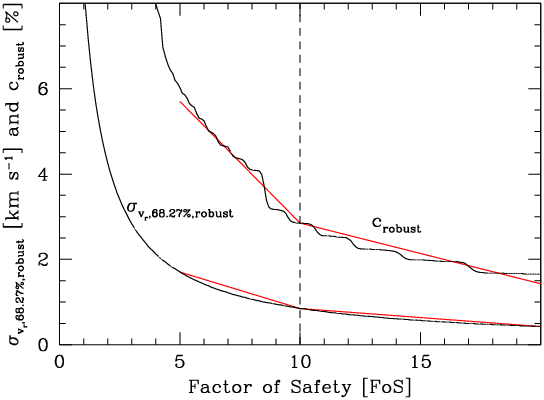}
  \color{black}
  \caption{\color{black}Dependence, for HIP$57367$ (see Table~\ref{tab:worst_confidence_level_robust}), of $\sigma_{v_r}$ (required to reach \color{black} $c_{\rm robust} = 68.27\%$\color{black}) and the robust \color{black} confidence level $c_{\rm robust}$ \color{black} on the Factor of Safety (FoS), i.e., the minimum factor between the predicted HTPM standard error and the perspective-acceleration-induced HTPM proper-motion error caused by radial-velocity errors. The default FoS value adopted in this study is $10$. The object is representative for a nearby star with a well-determined parallax. The straight, red lines indicate linear scaling relations starting from the default value ${\rm FoS} = 10$.\color{black}}\label{fig:FoS}
\end{figure}

For both the Gaussian and the robust criteria, we adopt, somewhat arbitrarily, the rule that the perspective-acceleration-induced HTPM proper-motion error caused by radial-velocity errors shall be an order of magnitude smaller than the predicted standard error of the HTPM proper motion itself (Section~\ref{sec:threshold}). The adopted Factor of Safety (FoS) is hence $10$. Some readers may find that this 'rule' is too stringent. Unfortunately, there is no easy (linear) way to scale our results if the reader wants to adopt a different value for the FoS. Clearly, the value of $\Sigma$ in each Monte-Carlo run is linearly proportional to the FoS (see, for instance, Figure~\ref{fig:schematic}). However, the robust \color{black} confidence level $c_{\rm robust}$ \color{black} and the radial-velocity standard error $\sigma_{v_r}$ required to meet a certain value of \color{black} $c_{\rm robust}$ \color{black} do not linearly depend on $\Sigma$ but on the properties of the, in general asymmetric, distribution of the $N = 10,000$ values of $\Sigma$ resulting from the Monte-Carlo processing. To zeroth order, however, one can assume a linear relationship between $\sigma_{v_r}$ and $\Sigma$ and hence the adopted FoS, as is also apparent from the 'error-free' criterion $\sigma_{v_r} < \frac{1}{2} \cdot \Sigma$ derived \color{black} for $c = 68.27\%$ \color{black} in Section~\ref{sec:principle}. This is in particular a fair approximation for small variations around the default value (${\rm FoS} = 10$) in combination with nearby stars, which are most interesting because these are most sensitive to perspective acceleration. The Monte-Carlo distribution of $\Sigma$ values for these objects is generally well behaved, i.e., symmetric and with \color{black} $\sigma_\varpi \ll \varpi$ and hence $\sigma_\Sigma \ll \mu_\Sigma$ \color{black} (see, for instance, Figure~\ref{fig:HIP38_gauss}\color{black}; see also Section~\ref{sec:gaussian_criterion}\color{black}). Figure~\ref{fig:FoS} shows, as an example for the nearby star HIP$57367$ (see Table~\ref{tab:worst_confidence_level_robust}), how $\sigma_{v_r}$ (required for \color{black} $c_{\rm robust} = 68.27\%$\color{black}) and \color{black} $c_{\rm robust}$ \color{black} vary as function of the FoS. Linear scaling around the default ${\rm FoS} = 10$ provides a decent approximation, at least over the range $10 / 2 = 5 < {\rm FoS} < 20 = 10 \cdot 2$.

\subsection{Number of Monte-Carlo simulations}\label{sec:number_of_runs}

\color{black}

\color{black}

To take measurement errors in the XHIP astrometry and literature radial velocities into account, we adopt a Monte-Carlo scheme in which we run a number $N$ of Monte-Carlo simulations for each star in which we randomly vary the astrometric and spectroscopic data within their respective error bars (see Section~\ref{sec:measurement_errors}). Clearly, the higher the value of $N$ is, the more reliable the results are. We adopted $N = 10,000$ as a practical compromise, resulting in an acceptable, typical run time of $\sim$$1$~s per star as well as smooth distributions of $\Sigma$ (see, for instance, Figure~\ref{fig:HIP38_gauss} or \ref{fig:HIP8_tail}). To investigate the repeatability and hence reliability of our robust \color{black} confidence levels $c_{\rm robust}$ \color{black} and radial-velocity errors $\sigma_{v_r}$, we have repeated the entire processing with $N = 10,000$ runs $100$ times for the \color{black} $206$ \color{black} stars with \color{black} $c_{\rm robust} < 68.27\%$ \color{black} (Section~\ref{sec:application_results}) and find that the typical variation of the \color{black} confidence level \color{black} and the radial-velocity error, quantified by the standard deviation divided by the average of the distribution containing the $100$ results, is less than \color{black} $0.1$\% \color{black} and \color{black} $0.2$\%\color{black}, respectively; the maximum variation among the \color{black} $206$ \color{black} objects \color{black} is found for HIP$107711$ and \color{black} amounts to \color{black} $0.5$\% and $0.6$\%, \color{black} respectively.

\subsection{High-priority and challenging stars}\label{sec:high_priority_and_challenging_stars}

Table~\ref{tab:worst_confidence_level_robust} shows the ten stars with the lowest robust \color{black} confidence level\color{black}. These stars are the highest-priority targets for spectroscopic follow-up. Nine of the ten entries do have a literature radial velocity but one which is insufficiently precise. The highest-priority object (HIP$57367$) does not yet have a spectroscopic radial velocity and needs a measurement with a standard error better than 1 km~s$^{-1}$. For this particular object, this challenge seems insurmountable since it is one of the $20$ white dwarfs with Hipparcos astrometry \citep{1997A&A...325.1055V}, objects for which it is notoriously difficult to obtain -- even low-precision -- spectroscopic radial velocities.

Table~\ref{tab:most_challenging_robust} shows the ten stars, among the subset of stars with unacceptably-low robust \color{black} confidence level ($c_{\rm robust} < 68.27\%$)\color{black}, with the smallest radial-velocity standard errors required to raise the robust \color{black} confidence level to $c_{\rm robust} = 68.27\%$\color{black}. Since \color{black} $c_{\rm robust} < 68.27\%$\color{black}, these stars do clearly need spectroscopic follow-up. However, the radial-velocity standard errors reach values as small as $0.04$~km~s$^{-1}$, which is a real challenge, not only in terms of the required signal-to-noise ratio of the spectroscopic data but also in view of the definition of the radial-velocity zero-point at this level of precision \citep{2010A&A...524A..10C} as well as potential systematic errors in the radial velocities, both with instrumental origin and with astrophysical causes such as radial-velocity differences between various absorption lines etc. (see \citealt{2003A&A...401.1185L} for a detailed discussion of this and other effects).

\color{black}

\color{black}

\begin{table*}[t]
\begin{center}
\caption{The ten stars with the lowest robust \color{black} confidence level $c_{\rm robust}$ \color{black} in the XHIP catalogue. The column with header \color{black} '$\sigma_{v_r}$, $c_{\rm robust} = 68.27\%$' \color{black} denotes the radial-velocity standard error that should be targeted in the spectroscopic follow-up to raise the \color{black} confidence level \color{black} of these stars to \color{black} $c_{\rm robust} = 68.27\%$\color{black}. The table contains two white dwarfs: HIP$57367$ (L$145-141$, type DQ, i.e., with carbon absorption features) and HIP$3829$ (van Maanen $2$, type DZ, i.e., with metal absorption features).}\label{tab:worst_confidence_level_robust}
\color{black}
\begin{tabular}[h]{rrrrcrrrrr}
\hline\hline
\\[-8pt]
\multicolumn{1}{c}{HIP} & \multicolumn{1}{c}{$v_r$} & \multicolumn{1}{c}{$\sigma_{v_r}$} & \multicolumn{1}{c}{\color{black}$c_{\rm robust}$\color{black}} & \multicolumn{1}{c}{\color{black}$\sigma_{v_r}$, $c_{\rm robust} = 68.27\%$\color{black}} & \multicolumn{1}{c}{$\varpi$} & \multicolumn{1}{c}{$\sigma_{\varpi}$} & \multicolumn{1}{c}{$H\!p$} & \multicolumn{1}{c}{$\mu_{\alpha}$} & \multicolumn{1}{c}{$\mu_{\delta}$}\\
& \multicolumn{1}{c}{[km~s$^{-1}$]} & \multicolumn{1}{c}{[km~s$^{-1}$]} & \multicolumn{1}{c}{\color{black}[\%]\color{black}} & \multicolumn{1}{c}{[km~s$^{-1}$]} & \multicolumn{1}{c}{[mas]} & \multicolumn{1}{c}{[mas]} & \multicolumn{1}{c}{[mag]} & \multicolumn{1}{c}{[mas yr$^{-1}$]} & \multicolumn{1}{c}{[mas yr$^{-1}$]}\\
\hline
\\[-8pt]
$ 57367$ &        -- &      -- & $  2.85$ & $  0.85$ & $217.01$ & $ 2.40$ & $11.5851$ & $ 2664.98$ & $ -348.60$\\
$ 73182$ & $  35.63$ & $10.06$ & $  4.22$ & $  0.53$ & $168.77$ & $21.54$ & $ 8.1824$ & $  961.78$ & $-1677.83$\\
$ 86990$ & $-115.00$ & $21.00$ & $  4.70$ & $  1.24$ & $171.48$ & $ 2.31$ & $10.7574$ & $-1119.35$ & $-1352.81$\\
$ 86214$ & $ -60.00$ & $21.00$ & $  6.16$ & $  1.62$ & $196.90$ & $ 2.15$ & $10.8964$ & $ -708.98$ & $ -937.40$\\
$  3829$ & $ 263.00$ & $ 4.90$ & $ 12.90$ & $  0.80$ & $234.60$ & $ 5.90$ & $12.5592$ & $ 1236.90$ & $-2709.19$\\
$ 72511$ & $ -39.60$ & $10.00$ & $ 13.40$ & $  1.65$ & $235.24$ & $44.85$ & $11.8102$ & $-1389.70$ & $  135.76$\\
$113229$ & $  46.60$ & $10.00$ & $ 16.51$ & $  2.08$ & $116.07$ & $ 1.19$ & $10.3732$ & $-1027.76$ & $-1060.78$\\
$ 72509$ & $ -38.80$ & $10.00$ & $ 16.87$ & $  2.07$ & $214.67$ & $43.88$ & $12.3140$ & $-1416.49$ & $ -270.45$\\
$ 80018$ & $  46.60$ & $10.00$ & $ 18.40$ & $  2.33$ & $118.03$ & $ 2.52$ & $10.5964$ & $ -740.10$ & $  997.40$\\
$ 55042$ & $ -35.00$ & $10.00$ & $ 20.15$ & $  2.55$ & $ 78.91$ & $ 2.60$ & $11.6071$ & $-2466.98$ & $ 1180.09$\\
\hline
\end{tabular}
\color{black}
\end{center}
\end{table*}

\begin{table*}[t]
\begin{center}
\caption{The ten stars, among the subset of stars with unacceptably-low robust \color{black} confidence level ($c_{\rm robust} < 68.27\%$)\color{black}, with the most stringent radial-velocity-error requirements $\sigma_{v_r}$ needed to reach \color{black} $c_{\rm robust} = 68.27\%$\color{black}.}\label{tab:most_challenging_robust}
\color{black}
\begin{tabular}[h]{rrrrcrrrrr}
\hline\hline
\\[-8pt]
\multicolumn{1}{c}{HIP} & \multicolumn{1}{c}{$v_r$} & \multicolumn{1}{c}{$\sigma_{v_r}$} & \multicolumn{1}{c}{\color{black}$c_{\rm robust}$\color{black}} & \multicolumn{1}{c}{\color{black}$\sigma_{v_r}$, $c_{\rm robust} = 68.27\%$\color{black}} & \multicolumn{1}{c}{$\varpi$} & \multicolumn{1}{c}{$\sigma_{\varpi}$} & \multicolumn{1}{c}{$H\!p$} & \multicolumn{1}{c}{$\mu_{\alpha}$} & \multicolumn{1}{c}{$\mu_{\delta}$}\\
& \multicolumn{1}{c}{[km~s$^{-1}$]} & \multicolumn{1}{c}{[km~s$^{-1}$]} & \multicolumn{1}{c}{\color{black}[\%]\color{black}} & \multicolumn{1}{c}{[km~s$^{-1}$]} & \multicolumn{1}{c}{[mas]} & \multicolumn{1}{c}{[mas]} & \multicolumn{1}{c}{[mag]} & \multicolumn{1}{c}{[mas yr$^{-1}$]} & \multicolumn{1}{c}{[mas yr$^{-1}$]}\\
\hline
\\[-8pt]
$ 87937$ & $-110.51$ & $ 0.10$ & $ 27.84$ & $  0.04$ & $548.31$ & $ 1.51$ & $ 9.4901$ & $ -798.80$ & $10277.30$\\
$ 54035$ & $ -84.69$ & $ 0.10$ & $ 51.74$ & $  0.07$ & $392.64$ & $ 0.67$ & $ 7.5062$ & $ -577.00$ & $-4761.80$\\
$114046$ & $   8.81$ & $ 0.10$ & $ 59.77$ & $  0.08$ & $305.26$ & $ 0.70$ & $ 7.4182$ & $ 6768.20$ & $ 1327.52$\\
$ 70890$ & $ -22.40$ & $ 0.50$ & $ 20.59$ & $  0.13$ & $772.33$ & $ 2.60$ & $10.7613$ & $-3775.64$ & $  768.16$\\
$   439$ & $  25.38$ & $ 0.22$ & $ 53.99$ & $  0.16$ & $230.42$ & $ 0.90$ & $ 8.6181$ & $ 5635.74$ & $-2338.18$\\
$ 73182$ & $  35.63$ & $10.06$ & $  4.22$ & $  0.53$ & $168.77$ & $21.54$ & $ 8.1824$ & $  961.78$ & $-1677.83$\\
$ 84140$ & $ -30.10$ & $ 1.10$ & $ 52.48$ & $  0.79$ & $158.17$ & $ 5.02$ & $ 9.3754$ & $  252.60$ & $-1571.80$\\
$ 10138$ & $  57.00$ & $ 0.80$ & $ 67.40$ & $  0.79$ & $ 92.74$ & $ 0.32$ & $ 6.2571$ & $ 2150.30$ & $  673.20$\\
$  3829$ & $ 263.00$ & $ 4.90$ & $ 12.90$ & $  0.80$ & $234.60$ & $ 5.90$ & $12.5592$ & $ 1236.90$ & $-2709.19$\\
$ 57367$ &        -- &      -- & $  2.85$ & $  0.85$ & $217.01$ & $ 2.40$ & $11.5851$ & $ 2664.98$ & $ -348.60$\\
\hline
\end{tabular}
\color{black}
\end{center}
\end{table*}

Table~2 in \citet{1999A&A...348.1040D} shows the top-$39$ of stars in the \citet{1995yCat.5070....0G} preliminary third catalogue of nearby stars ranked according to the magnitude of the perspective acceleration (which is propertional to $\varpi \cdot \mu$). Similarly, Table~1.2.3 in \citet{1997ESASP1200.....P} shows the top-21 of stars in the Hipparcos catalogue for which the magnitude of the perspective acceleration is significant enough to have been taken into account in the Hipparcos data processing (the accumulated effect on position is proportional to $\varpi \cdot \mu \cdot |v_r|$). On the contrary, the top-10 Tables~\ref{tab:worst_confidence_level_robust} and \ref{tab:most_challenging_robust} have been constructed based on the sensitivity of the perspective acceleration to radial-velocity errors (Section~\ref{sec:sensitivity_derivation}) and the associated \color{black} confidence level \color{black} of the available literature radial velocity. Hence, although there is a significant overlap of stars between the various tables, they are understandably not identical.

\subsection{Object-by-object analyses and other literature sources}

In general, and in particular for the most interesting, delicate, or border cases, it will be useful to perform a more in-depth literature search for and study of radial velocities and other available data before embarking on ground-based spectroscopy. For instance, we found a SIMBAD note on the Hipparcos catalogue \citep[CDS catalogue I/239]{1997ESASP1200.....P,1997yCat.1239....0E} for \color{black} HIP$114110$ ($c_{\rm robust} = 71.23\%$) and HIP$114176$ ($c_{\rm robust} = 60.00\%$) \color{black} that they are non-existing objects: {\it "HIP$114110$ (observed with HIP$14113$\footnote{This is a typo and must be HIP$114113$.}) and HIP$114176$ (observed with HIP$114177$) are noted as probable measurements of scattered light from a nearby bright star. The non-reality of $114110$ and $114176$ (traced to fictitious entries in the WDS and INCA) has been confirmed by MMT observations reported by D. Latham (private communication, 8 May 1998), and confirmed by inspection of the DSS [J.L. Falin, 12 May 1998]"}. In addition, the completeness and coverage level of the XHIP radial-velocity compilation is not known. We did query SIMBAD as well as the \color{black} Geneva-Copenhagen-Survey (GCS, CORAVEL) database \citep[CDS catalogue V/117]{2004A&A...418..989N} and the \color{black} RAVE database \citep[CDS catalogue III/265, with $77,461$ entries with a mean precision of $2.3$~km~s~$^{-1}$]{2011AJ....141..187S} for radial velocities for the \color{black} $109$ \color{black} stars without XHIP radial velocity and with confidence level below $68.27\%$ but did not find new data. Unfortunately, the treasure contained in the \color{black} full \color{black} CORAVEL database ($45,263$ late-type Hipparcos stars with precisions below $1$~km~s~$^{-1}$), the public release of which was announced in \citet{1997ESASP.402..693U} to be before the turn of the previous millennium, remains a mystery to date. \color{black} All in all, dedicated studies for individual objects might \color{black} pay off by reducing the needs for spectroscopic follow-up. 

\subsection{Urgency of the spectroscopic follow-up}

\citet{FM-040} already acknowledges that, since the perspective-acceleration-induced proper-motion error can be calculated as function of radial velocity, a factor -- effectively the sensitivity coefficients $C_\alpha$ and $C_\delta$ from Equations~(\ref{eq:Calpha})--(\ref{eq:Cdelta}) in Section~\ref{sec:sensitivity_derivation} -- can be published to correct the HTPM proper motion for a particular star {\it a posteriori} when $v_r$ becomes known or when a more precise $v_r$ becomes available. Therefore, both the reference radial velocity $v_r$ and parallax $\varpi$ used in the HTPM derivation will be published together with the proper-motion values themselves. This means that the spectroscopic follow-up identified in this paper is not time-critical: the HTPM catalogue can and will be published in any case, even if not all required spectroscopic follow-up has been completed. Of course, the implication for stars without the required radial-velocity knowledge will be that their HTPM proper motions will include a (potentially) significant perspective-acceleration-induced error.

\section{Conclusions}\label{sec:conclusion}

We have conducted a study of the requirements for the availability of radial velocities for the Hundred-Thousand-Proper-Motion (HTPM) project \citep{FM-040}. This unique project will combine Hipparcos astrometry from $1991.25$ with early-release Gaia astrometry ($\sim$$2014.5$) to derive long-time-baseline and hence precise proper motions. For the nearest, fast-moving stars, the perspective acceleration of the objects on the sky requires the presence of radial velocities for the derivation of the proper motions. We have quantitatively determined, for each star in the Hipparcos catalogue, the precision of the radial velocity that is required to ensure that the perspective-acceleration-induced error in the HTPM proper motion caused by the radial-velocity error is negligible. Our method takes the Hipparcos measurement errors into account and allows the user to specify his/her own prefered confidence level (e.g., $68.27\%$, $95.45\%$, or $99.73\%$). The results are available in Table~\ref{tab:datafile} (Appendix~\ref{sec:datafile}), which is published electronically only. We have compared the radial-velocity-precision requirements to the set of $46,392$ radial velocities contained in the XHIP compilation catalogue \citep{2012AstL...38..331A} and find that, depending on the confidence level one wants to achieve, hundreds to thousands of stars require spectroscopic follow-up. The highest-priority targets are \color{black} $206$ \color{black} objects with a confidence level below $68.27\%$; \color{black} $97$ \color{black} of them have a known but insufficiently precise radial velocity while the remaining \color{black} $109$ \color{black} objects have no literature radial velocity in the XHIP compilation catalogue at all. The typical brightness of the objects requiring their radial velocity to be (re-)determined is $H\!p \approx 8$--$12$~mag and the radial-velocity precisions vary drastically, ranging from $0.04$~km~s$^{-1}$ for the most extreme case (HIP$87937$, also known as Barnard's star) to a few tens of km~s$^{-1}$. With only few exceptions, the spectral types are K and M; \color{black} $73$\% \color{black} of them are in the south. Gaia's Radial-Velocity Spectrometer (RVS; \citealp{2011EAS....45..181C}) will deliver radial velocities for all stars in the HTPM catalogue with Gaia-end-of-mission precisions below a few km~s~$^{-1}$ (and $\sim$$10$~km~s~$^{-1}$ for early-type stars; \citealp{2012Ap&SS.tmp...68D}); however, these performances require full calibration of the instrument and data and hence will most likely only be reached in the final Gaia data release, at which time the HTPM proper motions will be superseded by the Gaia proper motions. Fortunately, the spectroscopic follow-up is not time-critical in the sense that the HTPM catalogue will be published with information (sensitivity coefficients and reference parallax and radial velocity) to correct the proper motions {\it a posteriori} when (improved) radial velocities become available. We finally note that the spectroscopic follow-up requirements for the HTPM proper motions quantified in this work will be dwarfed by the requirements coming from the end-of-mission Gaia proper motions, to be released around $\sim$$2021$: for instance for the stars in the HTPM catalogue, for which the HTPM proper-motion \color{black} standard \color{black} errors are $30$--$190$~$\mu$as~yr$^{-1}$, the Gaia proper-motion standard errors reach the bright-star floor around $3$--$4~\mu$as~yr$^{-1}$ \citep{2012Ap&SS.tmp...68D}, which means that the spectroscopic requirements for the correction of perspective acceleration in the Gaia astrometry will be a factor $\sim$$10$--$50$ more demanding.

\begin{acknowledgements}

It is a pleasure to thank Mark Cropper for discussions about radial-velocity surveys \color{black} and the referee, Fran\c{c}ois Mignard, for his constructive criticism which helped to improve our statistical methodology. \color{black} This research has made use of the SIMBAD database and VizieR catalogue access tool, both operated at the Centre de Donn\'{e}es astronomiques de Strasbourg (CDS), and of NASA's Astrophysics Data System (ADS).

\end{acknowledgements}

\appendix

\section{The results data file}\label{sec:datafile}

Table~\ref{tab:datafile} describes the results data file, which is available electronically only.

\begin{table*}[t]
\caption{Description of the final-results data file. The table, which is only available electronically, covers the $117,955$ entries of the XHIP catalogue (Section~\ref{sec:XHIP}). These are all Hipparcos-catalogue entries with astrometry. Columns 1--7 refer to data extracted from the XHIP catalogue described in Section~\ref{sec:XHIP}. \color{black} Columns 8--10 summarise some key quantities of our method. \color{black} Columns \color{black} 11--14 \color{black} provide data from the $N = 10,000$ Monte-Carlo simulations that we ran for each star (Section~\ref{sec:measurement_errors}). Columns \color{black}15--18 \color{black} refer to the Gaussian criterion developed in Section~\ref{sec:gaussian_criterion}. Columns \color{black}19--22 \color{black} refer to the robust criterion, which is described in Section~\ref{sec:robust_criterion}.}
\label{tab:datafile}
\begin{center}
\color{black}
\begin{tabular}[h]{cccl}
\hline\hline
Column & Value & Unit & Explanation\\
\hline
 1 &HIP & -- & Hipparcos identifier\\
 2 &$v_r$ & km~s$^{-1}$ & Radial velocity in the XHIP catalogue (Section~\ref{sec:XHIP_radial_velocities})\\
 3 &$\sigma_{v_r}$ & km~s$^{-1}$ & Standard error of the radial velocity in the XHIP catalogue (Section~\ref{sec:XHIP_radial_velocities})\\
 4 &$Q_{v_r}$ & -- & Quality grade of XHIP radial velocity ($1 = {\rm A}$, $2 = {\rm B}$, $3 = {\rm C}$, $4 = {\rm D}$ -- Section~\ref{sec:XHIP_radial_velocities})\\
 5 &$\varpi$ & mas & Parallax from the XHIP catalogue (Section~\ref{sec:XHIP_astrometry})\\
 6 &$\sigma_{\varpi}$ & mas & Parallax standard error from the XHIP catalogue (Section~\ref{sec:XHIP_astrometry})\\
 7 &$H\!p$ & mag & Hipparcos broad-band magnitude\\
 & & & \\
 8 &$C$ & $\mu$as~yr$^{-1}$ per km~s$^{-1}$ & Sensitivity of the proper motion to radial velocity ($C \equiv {{\partial \mu} / {\partial v_r}}$; Section~\ref{sec:sensitivity_derivation})\\
 9 &$\Sigma$ & km~s$^{-1}$ & Value of $\Sigma$ for the 'error-free' astrometry from the XHIP catalogue (Section~\ref{sec:principle})\\
10 &Flag & -- & Flag indicating whether columns 8--9 refer to right ascension ($=1$) or declination ($=2$)\\
 & & & \\
11 &$\Sigma_{\rm median}$ & km~s$^{-1}$& Median value of the $N = 10,000$ Monte-Carlo $\Sigma$ values (Section~\ref{sec:measurement_errors})\\
12 &$\Sigma_{\rm smallest}$ & km~s$^{-1}$& Smallest value of $\Sigma$ among the $N = 10,000$ Monte-Carlo simulations\\
13 &$\mu_\Sigma$ & km~s$^{-1}$& Mean of the Gauss fit to the histogram of the $N = 10,000$ Monte-Carlo $\Sigma$ values\\
14 &$\sigma_\Sigma$ & km~s$^{-1}$& Standard deviation of the Gauss fit to the histogram of the $N = 10,000$ Monte-Carlo $\Sigma$ values\\
 & & & \\
15 &$c_{\rm Gauss}$ & \% & Gaussian confidence level $c_{\rm Gauss}$ (Section~\ref{sec:gaussian_criterion})\\
16 &$\sigma_{v_r, 68.27\%, \rm{Gauss}}$ & km~s$^{-1}$& Radial-velocity precision needed for a $c_{\rm Gauss}=68.27\%$ confidence level\\
17 &$\sigma_{v_r, 95.45\%, \rm{Gauss}}$ & km~s$^{-1}$& Radial-velocity precision needed for a $c_{\rm Gauss}=95.45\%$ confidence level\\
18 &$\sigma_{v_r, 99.73\%, \rm{Gauss}}$ & km~s$^{-1}$& Radial-velocity precision needed for a $c_{\rm Gauss}=99.73\%$ confidence level\\
 & & & \\
19 &$c_{\rm robust}$ & \% & Robust confidence level $c_{\rm robust}$ (Section~\ref{sec:robust_criterion})\\
20 &$\sigma_{v_r, 68.27\%, {\rm robust}}$ & km~s$^{-1}$& Radial-velocity precision needed for a $c_{\rm robust}=68.27\%$ confidence level\\
21 &$\sigma_{v_r, 95.45\%, {\rm robust}}$ & km~s$^{-1}$& Radial-velocity precision needed for a $c_{\rm robust}=95.45\%$ confidence level\\
22 &$\sigma_{v_r, 99.73\%, {\rm robust}}$ & km~s$^{-1}$& Radial-velocity precision needed for a $c_{\rm robust}=99.73\%$ confidence level\\
\hline
\end{tabular}
\color{black}
\end{center}
\color{black} Detailed notes:
\begin{itemize}
\item Columns 2--4: for stars without literature radial velocity, these columns list NaN (see Section~\ref{sec:XHIP_radial_velocities});
\item Columns 8--22: for the $11,171$ entries with insignificant parallax measurements in the XHIP catalogue, i.e., $3,920$ entries with $\varpi \leq 0$ and $7,251$ entries with $0 < \varpi / \sigma_\varpi \leq 1$, these columns list NaN (see Section~\ref{sec:measurement_errors});
\item Columns 16--18: if, for the given value of the confidence level $c_{\rm Gauss}$, the argument of the square root in Equation~(\ref{eq:sigma_vr_gauss}) is negative, these columns list NaN (see Section~\ref{sec:gaussian_criterion});
\item Columns 16--18 and 20-22: if the radial-velocity precision (standard error) to reach a certain confidence level exceeds $9999.99$~km~s$^{-1}$, a value of $9999.99$~km~s$^{-1}$ is listed.
\color{black}\end{itemize}
\end{table*}


\begin{thebibliography}{57}
\expandafter\ifx\csname natexlab\endcsname\relax\def\natexlab#1{#1}\fi

\bibitem[{{Anderson} \& {Francis}(2012)}]{2012AstL...38..331A}
{Anderson}, E. \& {Francis}, C. 2012, Astronomy Letters, 38, 331

\bibitem[{{Antoja} {et~al.}(2011){Antoja}, {Figueras}, {Romero-G{\'o}mez},
  {Pichardo}, {Valenzuela}, \& {Moreno}}]{2011MNRAS.418.1423A}
{Antoja}, T., {Figueras}, F., {Romero-G{\'o}mez}, M., {et~al.} 2011, \mnras,
  418, 1423

\bibitem[{{Arenou} {et~al.}(1995){Arenou}, {Lindegren}, {Froeschle}, {Gomez},
  {Turon}, {Perryman}, \& {Wielen}}]{1995A&A...304...52A}
{Arenou}, F., {Lindegren}, L., {Froeschle}, M., {et~al.} 1995, \aap, 304, 52

\bibitem[{{Arenou} \& {Luri}(1999)}]{1999ASPC..167...13A}
{Arenou}, F. \& {Luri}, X. 1999, in Astronomical Society of the Pacific
  Conference Series, Vol. 167, Harmonizing Cosmic Distance Scales in a
  Post-HIPPARCOS Era, ed. D.~{Egret} \& A.~{Heck}, 13--32

\bibitem[{{Bailer-Jones}(2010)}]{2010MNRAS.403...96B}
{Bailer-Jones}, C.~A.~L. 2010, \mnras, 403, 96

\bibitem[{{Crifo} {et~al.}(2010){Crifo}, {Jasniewicz}, {Soubiran}, {Katz},
  {Siebert}, {Veltz}, \& {Udry}}]{2010A&A...524A..10C}
{Crifo}, F., {Jasniewicz}, G., {Soubiran}, C., {et~al.} 2010, \aap, 524, A10

\bibitem[{{Cropper} \& {Katz}(2011)}]{2011EAS....45..181C}
{Cropper}, M. \& {Katz}, D. 2011, in EAS Publications Series, Vol.~45, EAS
  Publications Series, 181--188

\bibitem[{{de Bruijne}(1999)}]{1999MNRAS.310..585D}
{de Bruijne}, J.~H.~J. 1999, \mnras, 310, 585

\bibitem[{{de Bruijne}(2012)}]{2012Ap&SS.tmp...68D}
{de Bruijne}, J.~H.~J. 2012, \apss, 68

\bibitem[{{de Bruijne} {et~al.}(2001){de Bruijne}, {Hoogerwerf}, \& {de
  Zeeuw}}]{2001A&A...367..111D}
{de Bruijne}, J.~H.~J., {Hoogerwerf}, R., \& {de Zeeuw}, P.~T. 2001, \aap, 367,
  111

\bibitem[{{de Bruijne} {et~al.}(2010){de Bruijne}, {Kohley}, \&
  {Prusti}}]{2010SPIE.7731E..35D}
{de Bruijne}, J.~H.~J., {Kohley}, R., \& {Prusti}, T. 2010, in Society of
  Photo-Optical Instrumentation Engineers (SPIE) Conference Series, Vol. 7731,
  Society of Photo-Optical Instrumentation Engineers (SPIE) Conference Series

\bibitem[{{Delhaye}(1965)}]{1965gast.conf...61D}
{Delhaye}, J. 1965, in Galactic Structure, ed. {A.~Blaauw \& M.~Schmidt}, 61

\bibitem[{{Dravins} {et~al.}(1999){Dravins}, {Lindegren}, \&
  {Madsen}}]{1999A&A...348.1040D}
{Dravins}, D., {Lindegren}, L., \& {Madsen}, S. 1999, \aap, 348, 1040

\bibitem[{{Elias} {et~al.}(2006){Elias}, {Alfaro}, \&
  {Cabrera-Ca{\~n}o}}]{2006AJ....132.1052E}
{Elias}, F., {Alfaro}, E.~J., \& {Cabrera-Ca{\~n}o}, J. 2006, \aj, 132, 1052

\bibitem[{{ESA}(1997{\natexlab{a}})}]{1997ESASP1200.....P}
{ESA}, ed. 1997{\natexlab{a}}, ESA Special Publication, Vol. 1200, {The
  HIPPARCOS and TYCHO catalogues. Astrometric and photometric star catalogues
  derived from the ESA HIPPARCOS Space Astrometry Mission}

\bibitem[{{ESA}(1997{\natexlab{b}})}]{1997yCat.1239....0E}
{ESA}. 1997{\natexlab{b}}, VizieR Online Data Catalog, 1239, 0

\bibitem[{{Eyer} {et~al.}(2011){Eyer}, {Suveges}, {Dubath}, {Mowlavi}, {Greco},
  {Varadi}, {Evans}, \& {Bartholdi}}]{2011EAS....45..161E}
{Eyer}, L., {Suveges}, M., {Dubath}, P., {et~al.} 2011, in EAS Publications
  Series, Vol.~45, EAS Publications Series, 161--166

\bibitem[{{Gliese} \& {Jahreiss}(1995)}]{1995yCat.5070....0G}
{Gliese}, W. \& {Jahreiss}, H. 1995, VizieR Online Data Catalog, 5070, 0

\bibitem[{{G{\'o}mez} {et~al.}(2010){G{\'o}mez}, {Helmi}, {Brown}, \&
  {Li}}]{2010MNRAS.408..935G}
{G{\'o}mez}, F.~A., {Helmi}, A., {Brown}, A.~G.~A., \& {Li}, Y.-S. 2010,
  \mnras, 408, 935

\bibitem[{{Gradshteyn} \& {Ryzhik}(2007)}]{2007gr...book.....G}
{Gradshteyn}, I.~S. \& {Ryzhik}, M. 2007, {Tables of Integrals, Series, and
  Products (7$^{\rm th}$ edition)}, ed. A.~{Jeffrey} \& D.~{Zwillinger}
  (Academic Press)

\bibitem[{{H{\o}g} {et~al.}(2000{\natexlab{a}}){H{\o}g}, {Fabricius},
  {Makarov}, {Urban}, {Corbin}, {Wycoff}, {Bastian}, {Schwekendiek}, \&
  {Wicenec}}]{2000yCat.1259....0H}
{H{\o}g}, E., {Fabricius}, C., {Makarov}, V.~V., {et~al.} 2000{\natexlab{a}},
  VizieR Online Data Catalog, 1259, 0

\bibitem[{{H{\o}g} {et~al.}(2000{\natexlab{b}}){H{\o}g}, {Fabricius},
  {Makarov}, {Urban}, {Corbin}, {Wycoff}, {Bastian}, {Schwekendiek}, \&
  {Wicenec}}]{2000A&A...355L..27H}
{H{\o}g}, E., {Fabricius}, C., {Makarov}, V.~V., {et~al.} 2000{\natexlab{b}},
  \aap, 355, L27

\bibitem[{{Hoogerwerf} {et~al.}(2001){Hoogerwerf}, {de Bruijne}, \& {de
  Zeeuw}}]{2001A&A...365...49H}
{Hoogerwerf}, R., {de Bruijne}, J.~H.~J., \& {de Zeeuw}, P.~T. 2001, \aap, 365,
  49

\bibitem[{{Jordi} {et~al.}(2010){Jordi}, {Gebran}, {Carrasco}, {de Bruijne},
  {Voss}, {Fabricius}, {Knude}, {Vallenari}, {Kohley}, \&
  {Mora}}]{2010A&A...523A..48J}
{Jordi}, C., {Gebran}, M., {Carrasco}, J.~M., {et~al.} 2010, \aap, 523, A48

\bibitem[{{Joshi}(2007)}]{2007MNRAS.378..768J}
{Joshi}, Y.~C. 2007, \mnras, 378, 768

\bibitem[{{Katz} {et~al.}(2011){Katz}, {Cropper}, {Meynadier}, {Jean-Antoine},
  {Allende Prieto}, {Baker}, {Benson}, {Berthier}, {Bigot}, {Blomme},
  {Boudreault}, {Chemin}, {Crifo}, {Damerdji}, {David}, {David}, {Delle Luche},
  {Dolding}, {Fr{\'e}mat}, {Gerbier}, {Gerssen}, {G{\'o}mez}, {Gosset},
  {Guerrier}, {Guy}, {Hall}, {Hestroffer}, {Huckle}, {Jasniewicz}, {Ludwig},
  {Martayan}, {Morel}, {Nguyen}, {Ocvirk}, {Parr}, {Royer}, {Sartoretti},
  {Seabroke}, {Simon}, {Smith}, {Soubiran}, {Steinmetz}, {Th{\'e}venin},
  {Turon}, {Udry}, {Veltz}, \& {Viala}}]{2011EAS....45..189K}
{Katz}, D., {Cropper}, M., {Meynadier}, F., {et~al.} 2011, in EAS Publications
  Series, Vol.~45, EAS Publications Series, 189--194

\bibitem[{{Kordopatis} {et~al.}(2011){Kordopatis}, {Recio-Blanco}, {de
  Laverny}, {Bijaoui}, {Hill}, {Gilmore}, {Wyse}, \&
  {Ordenovic}}]{2011A&A...535A.106K}
{Kordopatis}, G., {Recio-Blanco}, A., {de Laverny}, P., {et~al.} 2011, \aap,
  535, A106

\bibitem[{{Kovalevsky}(1998)}]{1998A&A...340L..35K}
{Kovalevsky}, J. 1998, \aap, 340, L35

\bibitem[{{Lindegren} {et~al.}(2008){Lindegren}, {Babusiaux}, {Bailer-Jones},
  {Bastian}, {Brown}, {Cropper}, {H{\o}g}, {Jordi}, {Katz}, {van Leeuwen},
  {Luri}, {Mignard}, {de Bruijne}, \& {Prusti}}]{2008IAUS..248..217L}
{Lindegren}, L., {Babusiaux}, C., {Bailer-Jones}, C., {et~al.} 2008, in IAU
  Symposium, Vol. 248, IAU Symposium, ed. {W.~J.~Jin, I.~Platais, \&
  M.~A.~C.~Perryman}, 217--223

\bibitem[{{Lindegren} \& {Dravins}(2003)}]{2003A&A...401.1185L}
{Lindegren}, L. \& {Dravins}, D. 2003, \aap, 401, 1185

\bibitem[{{Lindegren} {et~al.}(2012){Lindegren}, {Lammers}, {Hobbs},
  {O'Mullane}, {Bastian}, \& {Hern{\'a}ndez}}]{2012A&A...538A..78L}
{Lindegren}, L., {Lammers}, U., {Hobbs}, D., {et~al.} 2012, \aap, 538, A78

\bibitem[{{Liu} {et~al.}(2011){Liu}, {Zhu}, \& {Hu}}]{2011RAA....11.1074L}
{Liu}, J.-C., {Zhu}, Z., \& {Hu}, B. 2011, Research in Astronomy and
  Astrophysics, 11, 1074

\bibitem[{{Madsen} {et~al.}(2002){Madsen}, {Dravins}, \&
  {Lindegren}}]{2002A&A...381..446M}
{Madsen}, S., {Dravins}, D., \& {Lindegren}, L. 2002, \aap, 381, 446

\bibitem[{{Mignard}(2009)}]{FM-040}
{Mignard}, F. 2009, Gaia Data Processing and Analysis Consortium (DPAC)
  technical note GAIA-C3-TN-OCA-FM-040, {T}he
  {H}undred-{T}housand-{P}roper-{M}otion project

\bibitem[{{Mignard} \& {Klioner}(2010)}]{2010IAUS..261..306M}
{Mignard}, F. \& {Klioner}, S.~A. 2010, in IAU Symposium, Vol. 261, IAU
  Symposium, ed. {S.~A.~Klioner, P.~K.~Seidelmann, \& M.~H.~Soffel}, 306--314

\bibitem[{{Mouret}(2011)}]{2011PhRvD..84l2001M}
{Mouret}, S. 2011, \prd, 84, 122001

\bibitem[{{Nordstr{\"o}m} {et~al.}(2004){Nordstr{\"o}m}, {Mayor}, {Andersen},
  {Holmberg}, {Pont}, {J{\o}rgensen}, {Olsen}, {Udry}, \&
  {Mowlavi}}]{2004A&A...418..989N}
{Nordstr{\"o}m}, B., {Mayor}, M., {Andersen}, J., {et~al.} 2004, \aap, 418, 989

\bibitem[{{Oort}(1927)}]{1927BAN.....3..275O}
{Oort}, J.~H. 1927, \bain, 3, 275

\bibitem[{{Perryman}(2009)}]{2009aaat.book.....P}
{Perryman}, M. 2009, {Astronomical Applications of Astrometry: Ten Years of
  Exploitation of the Hipparcos Satellite Data}, ed. {Perryman, M.} (Cambridge
  University Press)

\bibitem[{{Perryman} {et~al.}(2001){Perryman}, {de Boer}, {Gilmore}, {H{\o}g},
  {Lattanzi}, {Lindegren}, {Luri}, {Mignard}, {Pace}, \& {de
  Zeeuw}}]{2001A&A...369..339P}
{Perryman}, M.~A.~C., {de Boer}, K.~S., {Gilmore}, G., {et~al.} 2001, \aap,
  369, 339

\bibitem[{{Perryman} {et~al.}(1997){Perryman}, {Lindegren}, {Kovalevsky},
  {Ho{\o}g}, {Bastian}, {Bernacca}, {Cr{\'e}z{\'e}}, {Donati}, {Grenon},
  {Grewing}, {van Leeuwen}, {van der Marel}, {Mignard}, {Murray}, {Le Poole},
  {Schrijver}, {Turon}, {Arenou}, {Froeschl{\'e}}, \&
  {Petersen}}]{1997A&A...323L..49P}
{Perryman}, M.~A.~C., {Lindegren}, L., {Kovalevsky}, J., {et~al.} 1997, \aap,
  323, L49

\bibitem[{{Pourbaix}(2008)}]{2008IAUS..248...59P}
{Pourbaix}, D. 2008, in IAU Symposium, Vol. 248, IAU Symposium, ed. {W.~J.~Jin,
  I.~Platais, \& M.~A.~C.~Perryman}, 59--65

\bibitem[{{R{\"o}ser} {et~al.}(2010){R{\"o}ser}, {Demleitner}, \&
  {Schilbach}}]{2010AJ....139.2440R}
{R{\"o}ser}, S., {Demleitner}, M., \& {Schilbach}, E. 2010, \aj, 139, 2440

\bibitem[{{R{\"o}ser} {et~al.}(2008){R{\"o}ser}, {Schilbach}, {Schwan},
  {Kharchenko}, {Piskunov}, \& {Scholz}}]{2008A&A...488..401R}
{R{\"o}ser}, S., {Schilbach}, E., {Schwan}, H., {et~al.} 2008, \aap, 488, 401

\bibitem[{{Sch{\"o}nrich} \& {Binney}(2009)}]{2009MNRAS.399.1145S}
{Sch{\"o}nrich}, R. \& {Binney}, J. 2009, \mnras, 399, 1145

\bibitem[{{Sch{\"o}nrich} {et~al.}(2010){Sch{\"o}nrich}, {Binney}, \&
  {Dehnen}}]{2010MNRAS.403.1829S}
{Sch{\"o}nrich}, R., {Binney}, J., \& {Dehnen}, W. 2010, \mnras, 403, 1829

\bibitem[{{Schwarzschild}(1907)}]{1907..............S}
{Schwarzschild}, K. 1907, Nachrichten von der Gesellschaft der Wissenschaften
  zu G{\"o}ttingen, 614

\bibitem[{{Seeliger}(1900)}]{1900AN....154...65S}
{Seeliger}, H. 1900, Astronomische Nachrichten, 154, 65

\bibitem[{{Siebert} {et~al.}(2011){Siebert}, {Williams}, {Siviero}, {Reid},
  {Boeche}, {Steinmetz}, {Fulbright}, {Munari}, {Zwitter}, {Watson}, {Wyse},
  {de Jong}, {Enke}, {Anguiano}, {Burton}, {Cass}, {Fiegert}, {Hartley},
  {Ritter}, {Russel}, {Stupar}, {Bienaym{\'e}}, {Freeman}, {Gilmore}, {Grebel},
  {Helmi}, {Navarro}, {Binney}, {Bland-Hawthorn}, {Campbell}, {Famaey},
  {Gerhard}, {Gibson}, {Matijevi{\v c}}, {Parker}, {Seabroke}, {Sharma},
  {Smith}, \& {Wylie-de Boer}}]{2011AJ....141..187S}
{Siebert}, A., {Williams}, M.~E.~K., {Siviero}, A., {et~al.} 2011, \aj, 141,
  187

\bibitem[{{Sozzetti}(2011)}]{2011EAS....45..273S}
{Sozzetti}, A. 2011, in EAS Publications Series, Vol.~45, EAS Publications
  Series, 273--278

\bibitem[{{Tanga} {et~al.}(2008){Tanga}, {Hestroffer}, {Delb{\`o}}, {Frouard},
  {Mouret}, \& {Thuillot}}]{2008P&SS...56.1812T}
{Tanga}, P., {Hestroffer}, D., {Delb{\`o}}, M., {et~al.} 2008, \planss, 56,
  1812

\bibitem[{{Udry} {et~al.}(1997){Udry}, {Mayor}, {Andersen}, {Crifo}, {Grenon},
  {Imbert}, {Lindegren}, {Maurice}, {Nordstroem}, {Pernier}, {Prevot},
  {Traversa}, \& {Turon}}]{1997ESASP.402..693U}
{Udry}, S., {Mayor}, M., {Andersen}, J., {et~al.} 1997, in ESA Special
  Publication, Vol. 402, Hipparcos - Venice '97, ed. {R.~M.~Bonnet, E.~H{\o}g,
  P.~L.~Bernacca, L.~Emiliani, A.~Blaauw, C.~Turon, J.~Kovalevsky,
  L.~Lindegren, H.~Hassan, M.~Bouffard, B.~Strim, D.~Heger, M.~A.~C.~Perryman,
  \& L.~Woltjer}, 693--698

\bibitem[{{van Leeuwen}(2007)}]{2007ASSL..350.....V}
{van Leeuwen}, F., ed. 2007, Astrophysics and Space Science Library, Vol. 350,
  {Hipparcos, the New Reduction of the Raw Data}

\bibitem[{{van Leeuwen}(2008)}]{2008yCat.1311....0V}
{van Leeuwen}, F. 2008, VizieR Online Data Catalog, 1311, 0

\bibitem[{{Vauclair} {et~al.}(1997){Vauclair}, {Schmidt}, {Koester}, \&
  {Allard}}]{1997A&A...325.1055V}
{Vauclair}, G., {Schmidt}, H., {Koester}, D., \& {Allard}, N. 1997, \aap, 325,
  1055

\bibitem[{{Wilkinson} {et~al.}(2005){Wilkinson}, {Vallenari}, {Turon},
  {Munari}, {Katz}, {Bono}, {Cropper}, {Helmi}, {Robichon}, {Th{\'e}venin},
  {Vidrih}, {Zwitter}, {Arenou}, {Baylac}, {Bertelli}, {Bijaoui}, {Boschi},
  {Castelli}, {Crifo}, {David}, {Gomboc}, {G{\'o}mez}, {Haywood}, {Jauregi},
  {de Laverny}, {Lebreton}, {Marrese}, {Marsh}, {Mignot}, {Morin}, {Pasetto},
  {Perryman}, {Pr{\v s}a}, {Recio-Blanco}, {Royer}, {Sellier}, {Siviero},
  {Sordo}, {Soubiran}, {Tomasella}, \& {Viala}}]{2005MNRAS.359.1306W}
{Wilkinson}, M.~I., {Vallenari}, A., {Turon}, C., {et~al.} 2005, \mnras, 359,
  1306

\bibitem[{{Zacharias} {et~al.}(2010){Zacharias}, {Finch}, {Girard}, {Hambly},
  {Wycoff}, {Zacharias}, {Castillo}, {Corbin}, {DiVittorio}, {Dutta}, {Gaume},
  {Gauss}, {Germain}, {Hall}, {Hartkopf}, {Hsu}, {Holdenried}, {Makarov},
  {Martinez}, {Mason}, {Monet}, {Rafferty}, {Rhodes}, {Siemers}, {Smith},
  {Tilleman}, {Urban}, {Wieder}, {Winter}, \& {Young}}]{2010AJ....139.2184Z}
{Zacharias}, N., {Finch}, C., {Girard}, T., {et~al.} 2010, \aj, 139, 2184

\end{thebibliography}
\end{document}